\documentclass[aps,prx,reprint,showpacs,superscriptaddress,longbibliography]{revtex4-2}
\usepackage[compat=1.0.0]{tikz-feynman}

\usepackage{amssymb}
\usepackage{bm}
\usepackage{amsmath}
\usepackage{graphicx}
\usepackage{epstopdf}
\usepackage{subfigure}
\usepackage{natbib}
\usepackage{epsfig}
\usepackage{amsfonts}
\usepackage{braket}
\usepackage{mathrsfs}
\usepackage{soul}
\usepackage[toc,page,title,titletoc,header]{appendix}
\usepackage[colorlinks,linkcolor=blue,citecolor=blue,anchorcolor=blue]{hyperref}
\usepackage{dsfont,amsthm,amsbsy}
\usepackage{booktabs}

\begin{document}

\title{Robust fluctuating intertwined charge stripes in the Emery model}

\author{Rong Zhang}
\email{rozhang@stanford.edu}

\affiliation{Department of Applied Physics, Stanford University, Stanford, CA
94305, USA}
 \affiliation{Stanford Institute for Materials and Energy Sciences,
 SLAC National Accelerator Laboratory, 2575 Sand Hill Road, Menlo Park, CA 94025, USA}

\author{Sijia Zhao}

\affiliation{Department of Applied Physics, Stanford University, Stanford, CA
 94305, USA}
 \affiliation{Stanford Institute for Materials and Energy Sciences,
 SLAC National Accelerator Laboratory, 2575 Sand Hill Road, Menlo Park, CA 94025, USA}

\author{Hong-Chen Jiang}
 \affiliation{Stanford Institute for Materials and Energy Sciences,
 SLAC National Accelerator Laboratory, 2575 Sand Hill Road, Menlo Park, CA 94025, USA}

\author{Brian~Moritz}
 \affiliation{Stanford Institute for Materials and Energy Sciences,
 SLAC National Accelerator Laboratory, 2575 Sand Hill Road, Menlo Park, CA 94025, USA}

\author{Edwin W. Huang}
\affiliation{Department of Physics and Institute of Condensed Matter Theory,
University of Illinois at Urbana-Champaign, Urbana, IL 61801, USA.
}

\author{Thomas P. Devereaux}
\email{tpd@stanford.edu}
\affiliation{Department of Materials Science and Engineering, Stanford University, Stanford, CA
94305, USA}
 \affiliation{Stanford Institute for Materials and Energy Sciences,
 SLAC National Accelerator Laboratory, 2575 Sand Hill Road, Menlo Park, CA 94025, USA}
\affiliation{Geballe Laboratory for Advanced
Materials, Stanford University, CA 94305.}

\date{\today}
 
\begin{abstract}
\noindent The single-band Hubbard model is one of the most extensively studied models in condensed matter physics, giving rise to intertwined spin and charge stripes that coexist with, or lie in the vicinity of, superconductivity in the phase diagram. However, whether the low energy physics of the single-band Hubbard model is fully equivalent to the multi-band (multi-orbital) Emery model remains an unsettled question. While the intertwined stripes and nematicity have been studied in the single-band Hubbard model, a comprehensive picture in the Emery model is lacking. In this paper, we focus on the less investigated intertwined charge stripes using complementary density matrix renormalization group (DMRG) and determinant quantum Monte Carlo (DQMC) techniques. Our ground state DMRG confirms the presence of the oxygen-centered charge stripes at a reduced amplitude in the Emery model parameter regime widely used in the study of intertwined stripes. Close analysis of the oxygen orbital structure of the static charge correlation function from DQMC reveals the charge stripe pattern in real-space, showcasing stronger charge density modulation on $p$-orbitals pointing along ``the rivers of charge'', consistent with DMRG. For the parameter set with the largest fermion signs, we managed to reach a temperature where the system first demonstrated tendencies to form purely unidirectional spin and charge stripes, and the $B_{1g}$ component becomes dominant in the bond-charge nematic susceptibility. This observation correlates with the doping dependence of the kinetic energy anisotropy, suggesting a close relation between the nematicity and charge stripes in the Emery model.

\end{abstract}

\maketitle

\section{Introduction}

A long-standing problem in condensed matter physics is an understanding of intertwined orders and the way in which many candidate ground states emerge from simple models containing electron correlations. None has been more studied than the single band Hubbard model - simply containing electron kinetic energy of bandwidth $W=8t$, with $t$ the nearest neighbor hopping, and an on-site-only Coulomb interaction parametrized by the Hubbard $U$. While perturbative regions (both $U/W << 1$ for any doping, and $W/U<<1$ near half-filling) have enabled well-controlled studies via renormalization group and projected Hilbert space methods, respectively, intermediate values of $U/W$ have been the realm of numerical studies using many methods \cite{Arovas_2022}. Density-matrix renormalization group (DMRG) methods likely have
provided the most detailed window into the important ground-state ordering tendencies at short to intermediate length scales. 
In this region many candidate ground states have emerged - such as spin density waves, charge density waves, $d_{x^2-y^2}$ superconductivity, as well as intertwined variants with a superposition of these orders - all appear to have ground state energies that can be separated by one part in a thousand of the hopping $t$. If we make a rough estimate with photoemission on the cuprate materials - as a candidate system thought to be related to the Hubbard model - $t\sim250$ meV and the ground states are separated by only a few degrees Kelvin. This has provided a challenge to the field, where numerical methods have predicted that either intertwined stripes or superconductivity may prevail at the lowest temperatures \cite{Zaanen_memorial}.\\

Yet, the single-band Hubbard model is only a rough proxy for the cuprates. Viewed as charge-transfer insulators at half-filling from numerous experimental characterizations, early theoretical attempts to mimic the cuprates stemmed from multi-band Hubbard models involving copper and oxygen degrees of freedom. \cite{Emery1987} Foundational studies focused on whether a salient ``bare minimum model'' could be established, and whether this model contained separate oxygen and copper degrees of freedom. A key finding of the time, albeit still controversial today, is that a so-called Zhang-Rice singlet (ZRS) \cite{ZRS} - a combination of a hole on oxygen orbiting a hole on copper with a definite phase convention - would be sufficient enough to form a building block for broader models where the singlet could be treated as a single ``doped hole'' away from half-filling. Despite being well-known that the ZRS cannot be defined as a bound object over the entire Brillouin zone, it was reasonable to conclude that as far as the low-energy degrees of freedom are concerned, as long as a Fermi surface lay close to the anti-ferromagnetic zone boundary where the ZRS is well defined, a single-band Hubbard model may be sufficient.\\

In the years that followed, this issue has never been ``settled". Indeed, given the strong sensitivity of numerical studies to intertwined ground states, a fundamental question has lingered as to whether more realistic models for the cuprates - including oxygen degrees of freedom that are captured over the entire Brillouin zone - might show more definitive stability towards one of the numerous phases observed in single-band Hubbard model simulations at intermediate values of $U/W$. \\

Despite the larger Hilbert space dimensions, DMRG studies have already weighed in on this issue on snorter ladders. It was found on 2-leg \cite{Song2021, 2-leg-pdw, Yang2024} and 
4-leg ladders \cite{White_2015, threebandSpinStripe, Ponsioen2023, Jiang2023a} that bond-centered charge stripes are largely dominant across different hole dopings, resembling stripes in the single-band Hubbard model. These stripes have charge periods following that of half-filled stripes $\lambda\sim 1/(2p)$, with $p$ the doped hole concentration, and 
antiferromagnetic regions of Cu hole spins separated by domain walls, whereby the $p$ doped holes are strongly concentrated on oxygen orbitals. The ground state of the 2-leg ladders or cylinders is sensitive to boundary conditions and oxygen configurations \cite{Polat2025}, which affect the pairing between doped carriers. When $d$-wave superconducting order is not the ground state, the introduction of longer-ranged hoppings or attractive next-nearest neighbor interactions lead to a uniform $d$-wave superconductor \cite{Yang2024} or pair-density wave \cite{2-leg-pdw}, respectively. The 2-leg ladders were also shown to have \cite{Nishimoto2002, Nishimoto2009, Nishimoto2010} a loop current phase but outside of the realistic cuprate parameter range. \\

Issue with the fermion sign have not precluded studies using numerically exact quantum Monte Carlo methods \cite{Dopf, Kung, Peng2025a}. Studies on the tendencies to $d$-wave superconductivity indicate that the Emery model is promising \cite{Kuroki1996, Mai2021, 3-band-transport}, but are often inconclusive due to problematic fermion signs limiting consideration to high temperatures or weaker interactions than the physically relevant ones. Nevertheless, spin stripes on the copper $d$-orbitals have been found for temperatures onsetting around $T\sim J$, the antiferromagnetic exchange, and persist over a wide doping range\cite{threebandSpinStripe}. Further, tendencies to form charge stripes have been observed, possibly commensurate with the spin stripe on the oxygen orbitals\cite{PeizhiThreeband}. However, the signal was found to be weaker compared to the charge correlation functions between the copper $d$-orbitals. 
So far, an understanding of intertwined stripes in the Emery model is still at an early stage compared to that of the single-band Hubbard model, where DQMC simulations have captured fluctuating intertwined spin and charge stripes \cite{HubbardChargeStripe, HubbardSpinStripe} and strong nematic fluctuations \cite{HubbardNematic}, likely related to the charge stripes \cite{Nie2017}.\\

Given the greater degree of tunability present in the Emery model, where Coulomb interactions on both Cu and O can be varied and the degree of hybridization can be modified via the charge transfer energy, it is the goal of the present study to understand how spin and charge stripes - the predominant ordering tendencies in the single band Hubbard model - may be altered by parameter choices. Due again to the fermion sign, we limit focus to the accessible spin and charge density correlations, and examine in greater detail how oxygen charge order is manifest in different parameter regimes. Although the model is closer to the cuprates, and it is tempting to draw analogies to the behavior of clustered oxygen modulations seen in STM measurements at low temperatures \cite{Ye2023a}, our study is more motivated from the numerical point of view for the temperatures that we can access. \\

In this work, we comprehensively study the fluctuating intertwined charge stripe tendency on the oxygen orbitals using numerically exact unbiased DQMC simulations and ground state DMRG simulations. By comparison with the single-band Hubbard model, our results show a richer structure of charge and spin stripes in the Emery model, with noticeable modifications by changing the Coulomb interactions. We found that ground state spin and charge stripes are qualitatively similar across the different parameters, but the standard Emery model parameter has a stronger charge modulation amplitude and displays a clearer charge stripe commensurate with the spin stripe in static correlation functions in DQMC simulations. The densities on the $p$-orbitals oriented along the antiferromagnetic domain wall have the strongest modulation amplitude in DMRG, which corroborates our DQMC observation where linear response theory suggests that an impurity on $p$-orbitals in one direction could induce a charge modulation in the other direction. For parameters with the best fermion sign, we observed a temperature at which the $B_{1g}$ nematic susceptibility begins to dominate as the $B_{2g}$ susceptibility dies off, and the static correlation functions on a square lattice start to show the superposition of unidirectional stripes. Along with the doping dependence of the kinetic energy anisotropy on rectangular lattices, our results suggest a close relationship between charge stripes and nematicity. \\

The outline of the paper is as follows.
Sec.~\ref{sec:model} specifies the Emery model and three different parameter sets we choose to simulate, relevant to cuprates and used previously in the literature. Sec.~\ref{sec:method} defines correlation functions and susceptibilities. Sec.~\ref{sec:benchmark} presents the DMRG benchmark for the parameter set with smaller $U_{dd}$ and large $J$. Sec.~\ref{sec:stripe} interprets the three different components of the oxygen charge correlation functions and the doping dependence of the modulation wavevector. Sec.~\ref{sec: temperature} discusses the temperature evolution of the spin and charge stripe patterns, and the bond-charge nematicity, which indicates the tendency for kinetic energy anisotropy.  

\section{Model}\label{sec:model}
The three-band Hubbard model describes the CuO$_2$ plane in hole language. It consists of copper-centered $d_{x^2 - y^2}$ orbitals arranged on a square lattice, and oxygen-centered $p_x$ or $p_y$ orbitals between every pair of nearest-neighbor copper sites, forming $\sigma$-bonds with the $d_{x^2 - y^2}$ orbital. The Emery model captures the nearest-neighbor hopping $t_{pd}$ between $p$ and $d$ orbitals, and the hopping $t_{pp}$ between $p_x$ and $p_y$ orbitals. $U_i$ is the onsite Hubbard repulsion. Specifically, $U_i=U_{dd}$ for copper sites and $U_{pp}$ for oxygen sites, and $\Delta_{pd}$ represents the charge transfer energy between the copper and oxygen orbitals. The chemical potential $\mu$ is tuned to give various doping levels. The Hamiltonian consists of kinetic energy $K$, charge transfer energy, and the Hubbard interaction energy. In the grand canonical ensemble, the Hamiltonian is given by
\begin{align*}
\begin{split}
H &= K +\Delta_{pd}\sum_{i,\sigma}n_{i,\sigma}+\sum_i U_i n_{i,\uparrow}n_{i,\downarrow} - \mu\sum_{i,\sigma} n_{i,\sigma},\\
K &= -\sum_{i}t_{pd}(\xi^a_{i} + \xi^b_{i} + \xi^c_{i}  + \xi^d_{i})\\
-&\sum_{i}t_{pp}(\xi^e_{i} + \xi^f_{i} + \xi^g_{i}  + \xi^h_{i}),
\end{split}
\end{align*}
\noindent where the orbital sign convention is given by the following, as in \cite{Kung}
\begin{align*}
    \xi^{a}_{i} &= \sum_{\sigma}(p^{x\dagger}_{i,\sigma}d_{i,\sigma} + h.c.), \\
    \xi^{b}_{i} &= -\sum_{\sigma}(p^{y\dagger}_{i,\sigma}d_{i,\sigma} + h.c.), \\
    \xi^{c}_{i} &= -\sum_{\sigma}(p^{x\dagger}_{i-\hat{\mathbf{x}},\sigma}d_{i,\sigma} + h.c.), \\
    \xi^{d}_{i} &= \sum_{\sigma}(p^{y\dagger}_{i-\hat{\mathbf{y}},\sigma}d_{i,\sigma} + h.c.),    
\end{align*}
\begin{align*}
    \xi^{e}_{i} &= -\sum_{\sigma}(p^{x\dagger}_{i,\sigma}p_{i+\hat{x} - \hat{y},\sigma}^{y} + h.c.), \\
    \xi^{f}_{i} &= \sum_{\sigma}(p^{x\dagger}_{i,\sigma}p_{i-\hat{y},\sigma}^{y} + h.c.), \\
    \xi^{g}_{i} &= -\sum_{\sigma}(p^{x\dagger}_{i,\sigma}p_{i,\sigma}^{y} + h.c.), \\
    \xi^{h}_{i} &= \sum_{\sigma}(p^{x\dagger}_{i,\sigma}p_{i+\hat{\mathbf{x}},\sigma}^{y} + h.c.).    
\end{align*}\\
The summation index $i$ in the kinetic energy $K$ runs through all unit cells; in the charge transfer and chemical potential terms, it runs through all sites on the Lieb lattice where $n_{i,\sigma}$ is the local density operator for spin $\sigma$ at each site. A standard parameter set for the Emery model, nominally describing cuprates, is $t_{pd} = 1.13, t_{pp} = 0.49, \Delta_{pd} = 3.24, U_{dd} = 8.5$, and $U_{pp} = 4.1$ or $0.0$, each in units of eV \cite{Kung}. An alternative set of parameters used in DQMC studies of spin stripes \cite{threebandSpinStripe, PeizhiThreeband} has $U_{dd} = 6~e$V, $U_{pp} = 0$, $\Delta_{pd} = 3~e$V, and is otherwise the same as the standard parameters. Here, $U_{pp} = 4.1$~eV, $U_{pp} = 0$, and $U_{dd} = 6$ eV distinguish the three parameter sets we study in this work.\\

Presumably, the effective antiferromagnetic coupling $J$ is an important energy scale for stripe physics and $U_{dd} = 6$~eV has been used in studies of spin stripes due to the larger $J$. If we take the peak position of the spin dynamical structure factor $S(\mathbf{q} = (0,\pi),\omega)$ from DQMC simulations at the lowest accessible temperature as an estimate for $2J$ for each parameter set, the standard parameter set would have $2J_{U_{pp}=4.1} \sim 0.279~e$V \cite{3-band-transport} at a lowest accessible temperature $\beta = 6.5~e$V$^{-1}$, $T/J_{U_{pp} = 4.1} \sim 1.1$, the same order of magnitude as the antiferromagnetic exchange, The other set has a much larger exchange coupling $2J_{U_{dd} = 6} \sim 0.576~e$V \cite{threebandSpinStripe}, and the fermion sign is also better, allowing for simulations to significantly lower temperatures, $\beta = 12~e$V$^{-1}$ $T/J_{U_{dd} = 6} \sim 0.29$, well below the exchange interaction scale. \\

\section{Method}\label{sec:method}
\subsection{Charge and spin correlation and susceptibilities}

At finite temperatures, in linear response theory, the charge modulation induced by a perturbation on the local chemical potential is given by the connected static charge correlation function, and the spin modulation induced by a local magnetic field is given by the static spin correlation functions, which are defined as
\begin{align*}
    \chi_{c, ij}(\omega = 0) &= \int_0^{\beta} \left( \braket{n_{i}(\tau)n_j(0)} - \braket{n_i(\tau)}\braket{n_j(0)}\right)d\tau,\\
    \chi_{s, ij}(\omega = 0) &= \int_0^{\beta}  \braket{S^{z}_{i}(\tau)S^{z}_j(0)}d\tau,
\end{align*}
where $\beta$ is the inverse temperature, $i,j$ are different sites in real space, which could belong to different orbitals. We denote the real space correlation functions between two orbitals $\mu,\nu$ by $\chi_{c,ij}^{\mu\nu}(\omega = 0)$. \\

Well studied in previous work \cite{threebandSpinStripe, PeizhiThreeband}, the spin correlation is most significant on copper $d$-orbitals. For identifying the period of the spin stripes from real space, we focus on the staggered spin correlation on copper $d$-orbitals, which is defined as
\begin{align*}
    \chi_{s, ij}^*(\omega = 0) &=  (-1)^{(\mathbf{r_i} - \mathbf{r}_j)\cdot (\hat{x} + \hat{y})}\chi_{s, ij}(\omega = 0), 
\end{align*}
where the $i, j$ are $d$ orbital indices.\\

Real space correlation functions that possess $C_4$ symmetry, such as the spin correlations on $d$ orbital, or $p_x$-$p_y$ charge correlation functions, should show patterns corresponding to the momentum space superposition of the modulation wavevectors of stripes related by symmetry, as discussed in the Supplementary information of Ref.~\cite{threebandSpinStripe}. On a rectangular cluster that breaks $C_4$ symmetry, $\chi_s$ and $\chi_c^{xy}$ are biased towards stripes in a single direction, similar to the single-band counterpart. By symmetry arguments, $\chi_{c}^{xx}$ and $\chi_{c}^{yy}$ would show preferred stripe direction even on a square lattice, because it is the induced charge modulation by a perturbation that breaks $C_4$ symmetry. On a rectangular lattice stripe tendencies, as indicated by these correlation functions, can strengthen or compete with the bias from the lattice geometry, as observed in the following results.\\

The connected static spin and charge correlations are related to the spin and charge susceptibilities in momentum space via the Fourier transform.
\begin{align*}
    \chi_{s}(\mathbf{q},\omega = 0) = \frac{1}{N}
    \sum_{j}\exp{(-i\mathbf{q}\cdot (\mathbf{r}_j - \mathbf{r}_i))}\chi_{s,ij}(\omega = 0),\\
    \chi_{c}^{\mu\nu}(\mathbf{q},\omega = 0) = \frac{1}{N}
    \sum_{j}\exp{(-i\mathbf{q}\cdot (\mathbf{r}_j - \mathbf{r}_i))}\chi_{c,ij}(\omega = 0),
\end{align*}
where $\mu,\nu \in \set{d, x, y}$, which stand for the orbitals $d$, $p_x$, and $p_y$. $i$ denotes any sites with orbital $\nu$, and $j$ denotes the summation over all sites with orbital $\mu$. $\mathbf{r}_i, \mathbf{r}_j$ are vectors pointing to the real space coordinate of orbital $i, j$. When the system is unstable against a spin stripe or charge density wave order, $\chi_s(\mathbf{q})$ or $\chi_{c}^{\mu\nu}(\mathbf{q})$ diverges at the modulation wavevector $\mathbf{q^*}$. Nevertheless, at finite temperature, observing a double peak structure in momentum space or a stripe-like pattern in real space does not necessarily imply the presence of the other, especially when the real space correlation function shows a stripe pattern but decays quickly or when the momentum space double peak is not extremely sharp.\\

\subsection{Bond charge nematic susceptibility}
 Upon the formation of a unidirectional stripe, 
 charge transport along the two $d$-$p$ bond directions ($x$ or $y$) is expected to differ. Consider a small anisotropy in $t_{pd}$: $t_{pd}\to t_{pd} +  \delta t$ in the $x$ direction and $t_{pd}\to t_{pd} -  \delta t$ in the $y$ direction. The induced difference between the contributions from the horizontal and vertical $d-p$ bonds to the kinetic energy, which we will call the $B_{1g}$ kinetic energy anisotropy, is given by 
\begin{align*}
\begin{split}
    &K_{pd}^x - K_{pd}^y \\
    = &-(t_{pd}+\delta t)\sum_{i}(\xi^a_i + \xi^c_i) + (t_{pd}-\delta t)\sum_{i}(\xi_i^b + \xi_i^d),
\end{split}
\end{align*}
which is related to the bond-charge nematic susceptibility by
\begin{align*}
    \frac{\partial \langle K^x_{pd} - K^y_{pd}\rangle}{\partial \delta t}\Bigg|_{\delta t = 0} = t_{pd}\chi_{B_{1g}}(\omega = 0) + \delta t\langle K_{pd}\rangle /t_{pd}.
\end{align*}
The static $B_{1g}$ bond-charge nematic susceptibility $\chi_{B_{1g}}$ is given by 
\begin{align*}
\begin{split}
&\chi_{B_{1g}}(\omega = 0) = 
    \frac{\partial\langle\rho_{B_{1g}}\rangle}{\partial\delta t}\Bigg\rvert_{\delta t = 0} \\
    = &\int_{0}^{\beta}(\langle \rho_{B_{1g}}(\tau)\rho_{B_{1g}}(0)\rangle - \langle\rho_{B_{1g}}(\tau)\rangle\langle\rho_{B_{1g}}(0)\rangle) d\tau,
    \end{split}
\end{align*}
where the $\rho_{B_{1g}}$ operator is defined by
\begin{align*}
    \begin{split}
    {\rho}_{B_{1g}} =& \sum_i \xi^{a}_i - \xi^{b}_i + \xi^{c}_i - \xi^{d}_{i}.
    \end{split}
\end{align*}
 If a stripe forms along the $d$-$p$ direction, a small perturbation in $t_{pd}$ would induce a finite kinetic energy anisotropy, and $\chi_{B_{1g}}(\omega = 0)$ will diverge. Similarly, for a small anisotropy in $t_{pp}$, we can deduce a bond-charge nematic susceptibility with $B_{2g}$ symmetry, $\chi_{B_{2g}}$, where the $B_{2g}$ bond-charge operator given by 
\begin{align*}
\begin{split}
    \rho_{B_{2g}} &= \xi^e_i - \xi^{f}_i + \xi^{g}_i - \xi^{h}_i.\\
\end{split}
\end{align*}
 If a diagonal stripe forms, $\chi_{B_{2g}}(\omega =0)$ should diverge. These susceptibilities provide another way to characterize the tendency to spontaneoutly form a unidirectional stripe in a given direction. \\

\subsection{Numerical simulation}
 The spin correlation function $\langle S_i^zS_j^z\rangle$ and the charge density profile $\langle n_i^{\mu}\rangle$ are measured from density matrix renormalization group (DMRG) ground state calculations on $24\times4$ (hole-doping  $\delta \sim 1/6$ and $1/12$) or $32\times4$ (hole doping $\delta \sim 1/8$) clusters in the cylinder geometry with $d$ and $p_y$ orbitals terminating the cylinders at both ends, with no dangling $p_x$ orbitals, for the $U_{dd} = 6$~eV Emery model. Static spin and charge susceptibilities $\chi_{s,ij}(\omega = 0), \chi_{c,ij}^{\mu\nu}(\omega = 0)$ at finite temperatures are measured from determinant quantum Monte Carlo (DQMC) \cite{DQMC1, DQMC2, DQMC3} simulations on $16\times4$ and $8\times8$ clusters with periodic boundary conditions for the three model parameter sets, in order to track stripe periodicity. To examine the tendency to form unidirectional stripes, bond-charge nematic susceptibilities $\chi_{B_{1g}}(\omega = 0)$ and $\chi_{B_{2g}}(\omega = 0)$ are measured on $8\times 8$ clusters. \\

\section{Results and Discussion}

\begin{figure*}[hbt!]
    \centering
    \includegraphics[width=1.0\linewidth]{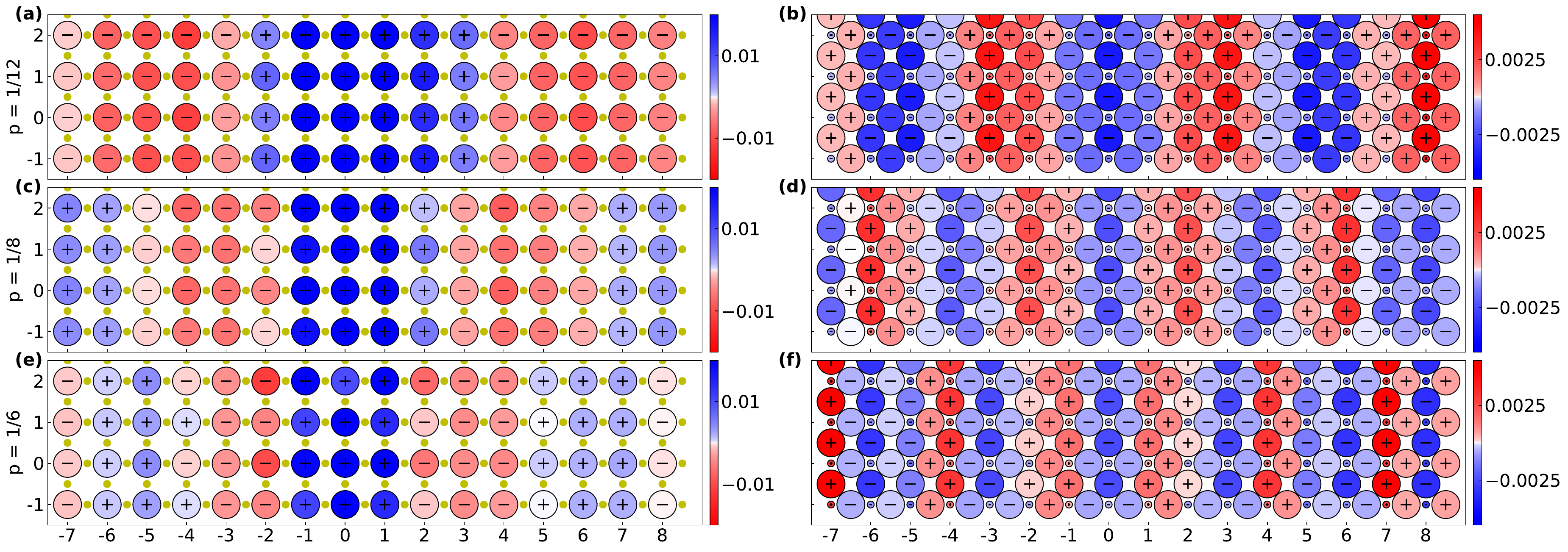}
    \caption{\textbf{Doping dependence of DMRG spin and charge stripes} \textbf{Left (a, c, e)}: staggered spin-spin correlations on copper
$d$ orbitals for increasing hole dopings, in reference to the $d$ orbital at (0, 0). \textbf{Right (b, d, f)}: charge density fluctuation profiles $\langle n_{i,\alpha}\rangle - \overline{\langle n_{\alpha}\rangle}, \alpha = d, p_x, p_y$, for each orbital plotted at their positions. The plotting convention and three-band parameters are the same as in Fig.~\ref{fig:doping-spin-locking}. The doping dependence of the spin and charge stripe period is consistent with that in Fig.~\ref{fig:doping-spin-locking}. The charge density fluctuation on $p_y$ is generally stronger than that on $p_x$. Calculations were performed on a $24 \times4$ or $32\times4$ cylinders, with the center $16 \times4$ rungs shown here.}
    \label{fig:dmrg}
\end{figure*}

\subsection{DMRG ground state benchmark}\label{sec:benchmark}
 A previous DQMC study on the charge density wave in the Emery model \cite{PeizhiThreeband} observed weak charge correlations on the oxygen orbitals and a stronger charge density wave tendency on $d$ orbitals, which is different from the existing ground-state Emery model calculations \cite{White_2015, threebandSpinStripe, Ponsioen2023}. To investigate whether the different Emery model parameters contribute to the distinction, we performed a ground-state DMRG simulation for model parameters as in Ref.~\cite{PeizhiThreeband}. \\

 Figure~\ref{fig:dmrg} shows the staggered spin correlation on copper (left) and spatial fluctuations of the charge density profile (right) from a DMRG calculation of the inner 16 rungs of a $24\times 4$ ($32\times 4$ for $1/8$ doping) cylinder at $U_{dd} = 6$~eV, $U_{pp} = 0$, $\Delta_{pd} = 3$~eV. The DMRG staggered spin correlations show a shorter spin stripe period with increasing hole doping. Charge accumulates near the AFM domain wall, while being depleted near the center of the cylinder. In panels (b, d, f), the charge densities on $d, p_x, p_y$ orbitals vary concomitantly with a common periodicity, locking to half of the spin stripe period. This charge modulation is strongest on the $p_y$ orbital and weakest on the $d$ orbitals. Although this parameter set features smaller $U_{dd}$, $U_{pp}$, and $\Delta_{pd}$ compared to Refs.~\cite{White_2015, threebandSpinStripe, Ponsioen2023}, the ground states remain qualitatively similar. Nevertheless, the charge modulation amplitude in our calculation is significantly weaker, as detailed by comparisons in Fig.~\ref{fig:DMRG}. \\

 Although at high temperatures \cite{PeizhiThreeband} the modulation wavevector $\mathbf{q}^*$ of $\chi_{c}^{dd}(\mathbf{q}, \omega = 0)$ does not follow the Yamada relation, it gives way to a charge density wave with a period locked to twice the spin stripe period in the ground state, which is also present on oxygen components of $\chi_{c}(\mathbf{q},\omega = 0)$ at finite temperature. This benchmark calculation demonstrates that there is likely no qualitative difference among the three parameter sets we studied, at least with respect to the static charge modulation on oxygen orbitals.\\

\begin{figure*}[hbt!]
    \centering
    \includegraphics[width=1.0\linewidth]{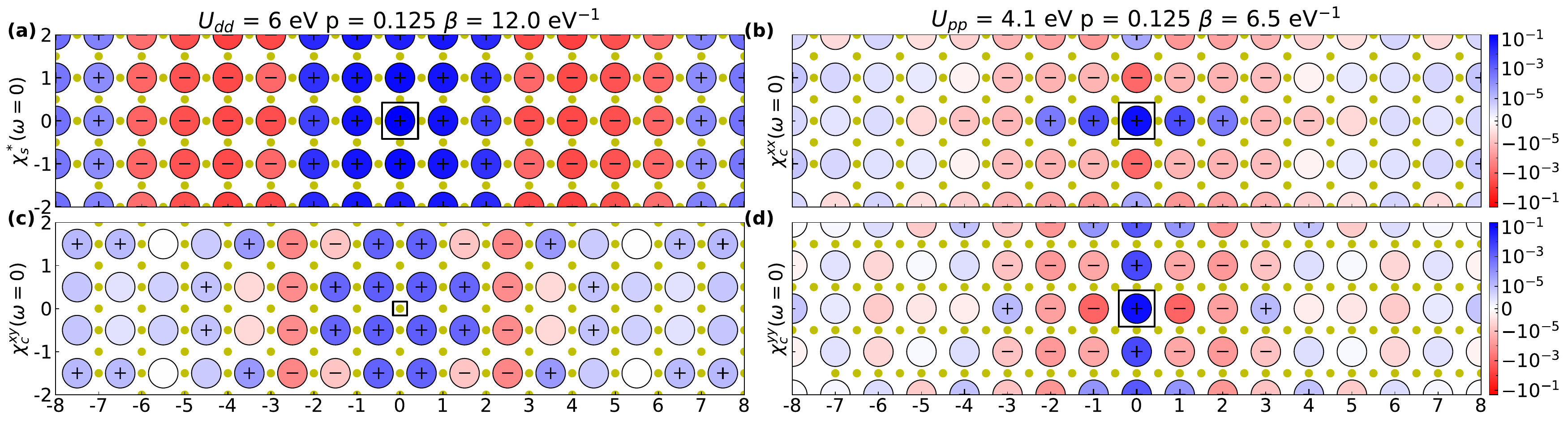}
    \caption{\textbf{Real space patterns of DQMC spin and charge correlation functions} \textbf{(a)} Staggered spin correlation function $\chi_{s,ij}^*(\omega = 0)$, showing fluctuating spin stripe domains staggered across the length of the cylinder and running along the vertical direction. \textbf{(b)} Static charge correlation function between $p_x$ orbitals $\chi_{c}^{xx}(\omega = 0)$, showing a tendency to form horizontal stripes due to the local perturbation on the $x$ orbital. \textbf{(c)} Static charge correlation function between $p_x$ and $p_y$ orbitals $\chi_{c}^{xy}(\omega = 0)$, showing a tendency to form stripes along the vertical direction due to the lattice geometry, where ``rivers of charge" would be pinned to the spin stripe domain walls. \textbf{(d)} Static charge correlation function between $p_y$ orbitals $\chi_{c}^{yy}(\omega = 0)$, showing a tendency to form stripes along the vertical direction due to local perturbation. Data from different model parameters are presented here for clarity of the patterns. Other data are presented in Figs.~\ref{fig:doping-spin-locking-x} through 
    \ref{fig:doping-spin-locking-yy-Upp}.}
    \label{fig:doping-spin-locking-U}
\end{figure*}

\subsection{Fluctuating charge stripe on $p$-orbitals}\label{sec:stripe}

\begin{figure*}[hbt!]
    \centering
    \includegraphics[width=1.0\linewidth]{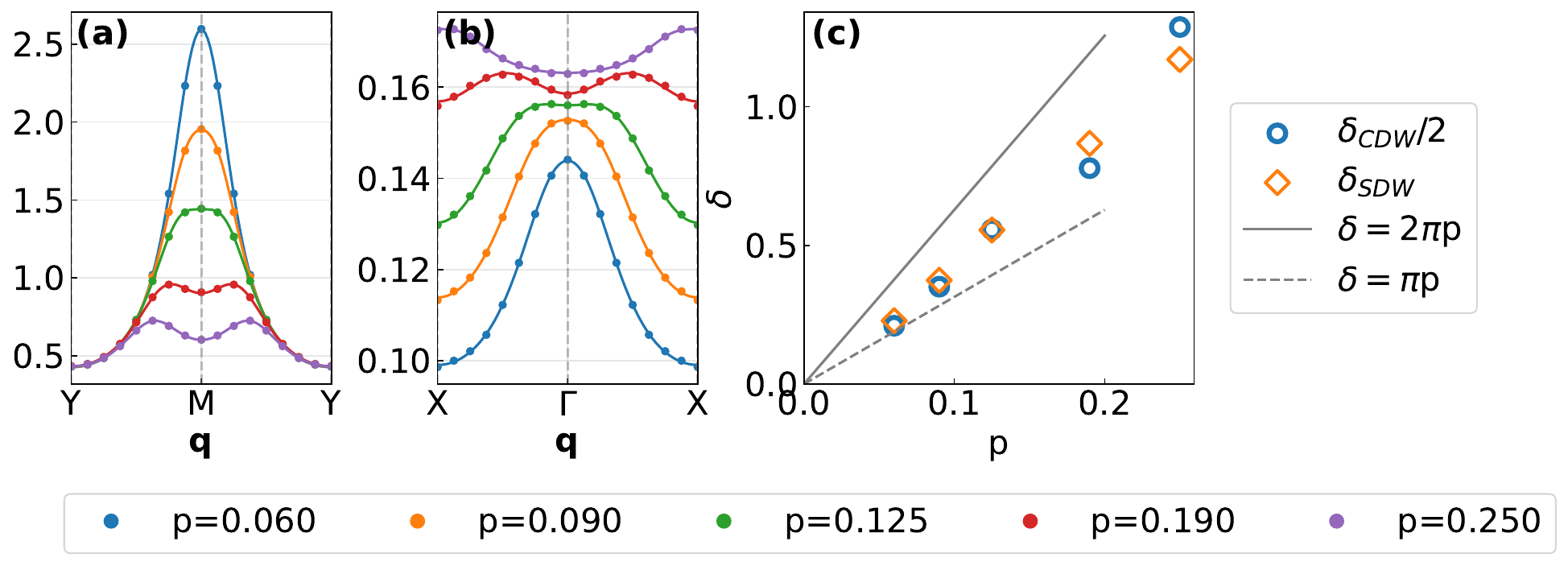}
    \caption{\textbf{Momentum space DQMC spin and charge correlation functions on a $16\times 4$ cluster with $U_{pp} = 0$, $U_{dd} = 8.5$~eV} \textbf{(a)} Static spin correlation function $\chi_{s}(\mathbf{q}, \omega = 0)$. \textbf{(b)} Fourier transform of the static oxygen charge correlation function with respect to $p_y$, $\chi_{c}^{yy}(\mathbf{q},\omega = 0) + \chi_{c}^{xy}(\mathbf{q},\omega = 0)$, In \textbf{(a)} and \textbf{(b)}, the markers represent the data, and solid lines are fitted periodic double Lorentzian functions as defined in Ref.~\cite{HubbardChargeStripe}. \textbf{(c)} The fitted modulation wavevectors of the spin density wave $\delta_{\textbf{SDW}}$ and half of the charge density wave $\delta_{\textbf{CDW}}/2$ at each doping $p$, showing a Yamada-like relation. Similar plots for other model parameters are shown in Figs.~\ref{fig:Upp41Yamada} and \ref{fig:Udd6Yamada}.}
    \label{fig:Upp0Yamada}
\end{figure*}

\begin{table*}[]
    \centering
    \begin{tabular}{|c|c|c|c|c|}
        \hline
        Model parameter & $\chi_{s,ij}$ correlation length & $\chi_{c}^{xx}(\mathbf{q})$ doping dependence & $\chi_{c,ij}^{yy}$ stripe pattern & $\chi_{c,ij}^{xx}$ stripe pattern \\
        \hline
        $U_{pp} = 4.1$ eV at $\beta = 6.5$ eV$^{-1}$ & decays fastest & Yes & Yes & Yes \\
        \hline
        $U_{pp} = 0$ at $\beta = 10.0$ eV & -- & Yes & Yes & not obvious \\
        \hline
        $U_{dd} = 6$ eV at $\beta = 12.0$ eV$^{-1}$ & decays slowest & not obvious & not clear & No\\
        \hline
    \end{tabular}
    \caption{Qualitative difference observed between the three sets of Emery model parameters}
    \label{tab:modelparameter}
\end{table*}
 From DQMC calculations on $16\times 4$ clusters, the spin stripe pattern is visible in $\chi_{s,ij}(\omega = 0)$ across all three parameter sets, as illustrated by the representative example in Fig.~\ref{fig:doping-spin-locking-U} (a). The correlation lengths decrease as $J$ decreases and as temperature increases (comprehensive data for all parameter sets are provided in the Supplemental Materials). Similarly, the cross-component $\chi_{c}^{xy}$, which is sensitive to the symmetry breaking of the cluster geometry (see Appendix for detail), exhibits clear stripe patterns across the parameters, with a representative case shown in Fig.~\ref{fig:doping-spin-locking-U} (c). Two columns of negative correlation functions are spaced by approximately the size of the antiferromagnetic domain shown in the spin correlation functions, indicating that the doped charges tend to either arrange in lines, or clusters. From the DMRG simulations, the charge rich or deficit regions are balanced in the ground state. Without a loss of generality, we refer to the negatively correlated region as charge accumulation.\\

For $U_{dd} = 8.5$~eV and $U_{pp} = 4.1$~eV, $\chi_{c}^{yy}$ and $\chi_{c}^{xx}$ display real-space charge stripe patterns along the vertical and horizontal directions, respectively (Fig.~\ref{fig:doping-spin-locking-U} (b, d)). This behavior indicates that a perturbation on the local potential of the $p_x$ orbital tends to induce a charge density wave along the $x$ direction, even when the cluster geometry favors a charge density wave along the $y$ direction. Furthermore, the pattern in $\chi_{c}^{xx}$ indicates that it is more strongly affected by the symmetry breaking of the local perturbation than by the cluster geometry. This $x$-direction pattern is less clear for $U_{dd} = 8.5$~eV and $U_{pp} = 0$~eV, and disappears for $U_{dd} = 6$~eV at the lowest accessible temperatures. From the DMRG benchmark, we know that the ground-state oxygen charge modulations are qualitatively similar across all three parameter sets. Therefore, the differing responses to local perturbations observed in DQMC at the lowest accessible temperatures likely result from a tradeoff between the intrinsic charge modulation strength and the fermion sign problem. \\

 In momentum space, the modulation vector $\delta_{SDW} = |\mathbf{q}^* - (\pi,\pi)|$ extracted from $\chi_s(\mathbf{q}, \omega = 0)$ along the $Y-M$ direction by fitting to a periodic double Lorentzian peaked at $\mathbf{q}^*$, increases with hole doping $p$. An example with $U_{pp} = 0$~eV is shown in Fig.~\ref{fig:Upp0Yamada} (a). For the charge susceptibility, the $d$-orbital component $\chi_{c}^{dd}(\omega = 0)$ shows two peaks whose separation decreases as $p$ increases, consistent with previous work \cite{PeizhiThreeband}. The $p$ orbital components $\chi_{c}^{xy}(\omega = 0)$, $\chi_{c}^{yy}(\omega = 0)$, and $\chi_{c}^{xx}(\omega = 0)$, exhibit two peaks whose separation increases with increasing $p$, or one single peak that broadens or remains unchanged if the two peaks are unresolvable. Generally, the peak separation is largest in $\chi_{c}^{yy}$ and smallest in $\chi_{c}^{xy}$. $\chi_c^{xx}$ shows no double peak structure along the $Y-M$ direction because a perturbation on the $p_x$ orbital favors a charge stripe along the horizontal direction. Table~\ref{tab:modelparameter} summarizes the qualitative differences in the static charge correlation functions, viewed from both momentum and real space, with supporting data provided in Fig.~\ref{fig:nnk},~\ref{fig:nnk-Upp0}, and \ref{fig:nnk-Upp41} from the Supplementary Materials. Overall, the charge modulation wavevector derived from the oxygen charge correlation functions is proportional to the spin stripe modulation wavevector. 

 Based on the preceding discussion, $\chi_c^{xx}$ is likely subject to a competing charge-ordering tendency distinct from that induced by the cluster geometry. Hence, it should be excluded when extracting the charge modulation wavevector for the vertical charge stripes present in $16\times 4$ clusters. Furthermore, because the ground-state charge modulation on the $d$ orbitals is significantly weaker than that on the $p$ orbitals, it is reasonable to focus on the oxygen charge modulation induced by a local perturbation of the $p_y$ charge transfer potential. This is represented by $\chi_c^{yy}(\mathbf{q},\omega = 0) + \chi_c^{xy}(\mathbf{q},\omega = 0)$ in Fig.~\ref{fig:Upp0Yamada} (b), which fits well to the periodic double Lorentzian function, similar to the charge correlation function of the single-band Hubbard model \cite{HubbardChargeStripe}. Figure~\ref{fig:Upp0Yamada} (c) shows that the modulation wavevector of the spin density wave is approximately half of that of the charge density wave, consistent with an intertwined spin and charge stripe scenario. The relation between the spin density wave wavevector and the hole doping shows that the stripe is partially filled at this temperature. For the other two parameter sets, the charge correlation function does not fit to periodic double Lorentzian peaks as cleanly, and the relation between $\delta_{\mathrm{CDW}}/2$ and $\delta_{\mathrm{SDW}}$ is less precise (Fig.~\ref{fig:Upp41Yamada}, \ref{fig:Udd6Yamada}), likely due to the high temperature or weaker modulation intensity, though they remain approximately proportional.\\

 In short, 
our DQMC and DMRG results on rectangular clusters demonstrate that the intertwined spin stripes on copper and charge stripes on oxygen are reachable by finite temperature DQMC, at least for Emery models with parameters that favor stronger charge modulation. Furthermore, the direction of the charge stripe, as shown in linear response, could be more sensitive to perturbation on the local potential of oxygen orbitals than the lattice geometry. However, because rectangular lattices break $C_4$ symmetry, these simulations do not probe the spontaneous tendency to form unidirectional stripes along the $d-p$ bond directions.\\

\subsection{Temperature evolution of stripes and bond-charge nematicity}\label{sec: temperature}

 Square lattices are essential for measuring the tendency towards unidirectional stripe formation in the $d-p$ bond direction via evolution of the stripe patterns in static correlation functions and $B_{1g}$ and $B_{2g}$ bond-charge nematic susceptibilities with decreasing temperature. In this geometry, the $p_x$-$p_y$ component of a static susceptibility is fully $C_4$ symmetric, whereas the $p_x$-$p_x$ and $p_y$-$p_y$ are related by $90^\circ$ rotation and each breaks $C_4$ symmetry. \\

 Ideally, the $C_4$-symmetric spin and charge density correlation functions showcase the following patterns when there is a tendency to form a unidirectional stripe, which breaks the $C_4$ symmetry in either the $B_{1g}$ or $B_{2g}$ mode. In the $B_{1g}$ mode, there is a tendency to form a stripe along the $d$-$p$ bond directions. The momentum-space superpositon of modulation wavevectors at $(\pi\pm\delta,\pi)$ and $(\pi,\pi\pm\delta)$ yields a diamond-like pattern in the staggered spin correlation on $d$ orbitals, and in $\chi_c^{xy}$ charge accumulation near the antiferromagnetic domain walls in the $\chi_{s,ij}^*$, which is near the corners of our square clusters. In the $B_{2g}$ mode, there is a tendency to form a stripe along the $p_x-p_y$ bond directions. The momentum-space superposition of modulation wavevectors at $(\pi\pm\delta, \pi\pm\delta)$ gives square-like antiferromagnetic domains, and the charge density should accumulate near the edges of the square clusters. These criteria identify the tendency to break $C_4$ symmetry in either a $B_{1g}$ or $B_{2g}$ pattern from the otherwise $C_4$ symmetric spin and charge density correlations. \\

\begin{figure*}[hbt]
    \centering
    \includegraphics[width=1.0\linewidth]{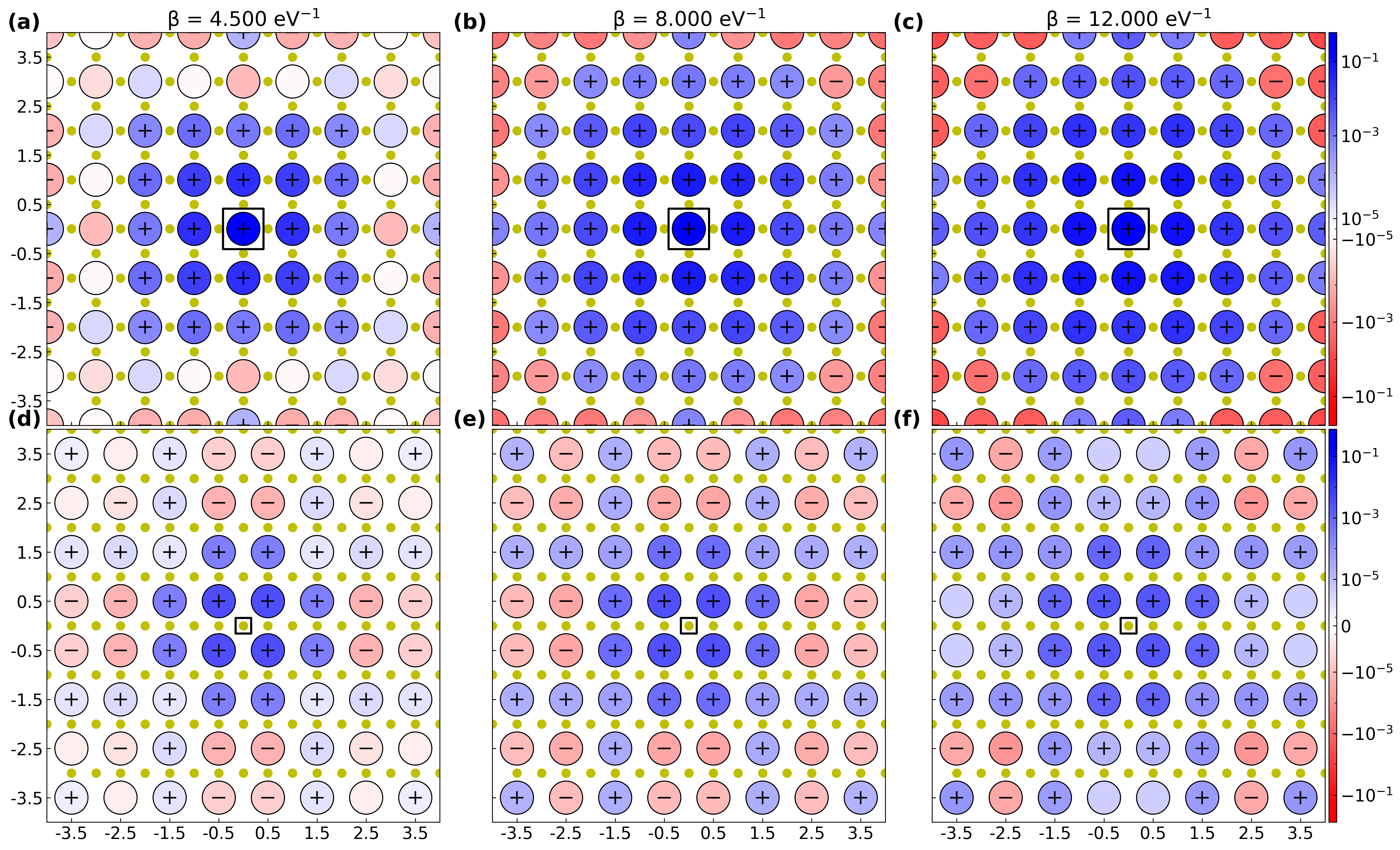}
    \caption{\textbf{Temperature evolution of staggered spin-spin (a, b, c) and connected $p_x-p_y$-orbital charge-charge correlation (d, e, f) from a DQMC simulation on $8\times 8$ system} The shape of the spin stripe domain in the correlation function gradually evolves from square-like at high temperatures in (a) towards the diamond-like shape in (e). The diamond-like domain wall is the momentum space superposition of horizontal and vertical spin stripes extending in the direction of the Cu-O bonds, while the square-like domain wall is the momentum space superposition of the spin stripes in the two diagonal directions. Meanwhile, at $\beta = 4.5~e$V$^{-1}$, the charge modulation tendency is strongest around $(0,\pm3), (\pm2,0)$ near the spin stripe domain walls. At $\beta = 12~e$V$^{-1}$, the charge modulation tendency is strongest near $(\pm 3, \pm 3)$, again near the domain walls. At $\beta = 8.0~e$V$^{-1}$, the AFM domain is a deformed diamond, while the charge modulation is a mixture of the high temperature and low temperature cases. The three-band parameters in this plot are $U_{dd} = 6~e$V, $U_{pp} = 0~e$V, at $p = 1/8$.} 
    \label{fig:temperature-locking}
\end{figure*}

Figure~\ref{fig:temperature-locking} displays the staggered spin (top row) and  $p_x$-$p_y$ charge (bottom row) correlation functions with temperature decreasing from left to right.
As seen in panels (a, b, c), lowering the temperature causes the local antiferromagnetic domain in the spin correlations to evolve from a square-like to a diamond-like shape. This evolution indicates a crossover from a tendency to break $B_{2g}$ to one which brakes $B_{1g}$ symmetry upon cooling. Concurrently, the charge correlations in panels (d, e, f) show that charge accumulation tendency weakens at the edge-aligned regions around $(0,\pm 3)$ and $(\pm 2, 0)$, while strengthening near the corner $(\pm 3, \pm 3)$. This intensity shift further corroborates the enhanced tendency to form unidirectional charge stripes along the $d$-$p$ bond directions. Overall, Fig.~\ref{fig:temperature-locking} demonstrates that the $\chi_{c,ij}^{xy}(\omega = 0)$ and $\chi_{s,ij}(\omega = 0)$ evolve concomitantly with doping and temperature, at least up to $T\simeq 0.2$~eV, above which the correlation functions are obscured by statistical noise. This tendency to form spin and charge stripes along the $x$ or $y$ directions is observed across other hole-doping levels and three-band parameter sets (see Fig.~\ref{fig:temperature-locking-U-n1.10_xy} and \ref{fig:temperature-locking-n1.20_xy}).\\

\begin{figure}
    \centering
    \includegraphics[width=1.0\linewidth]{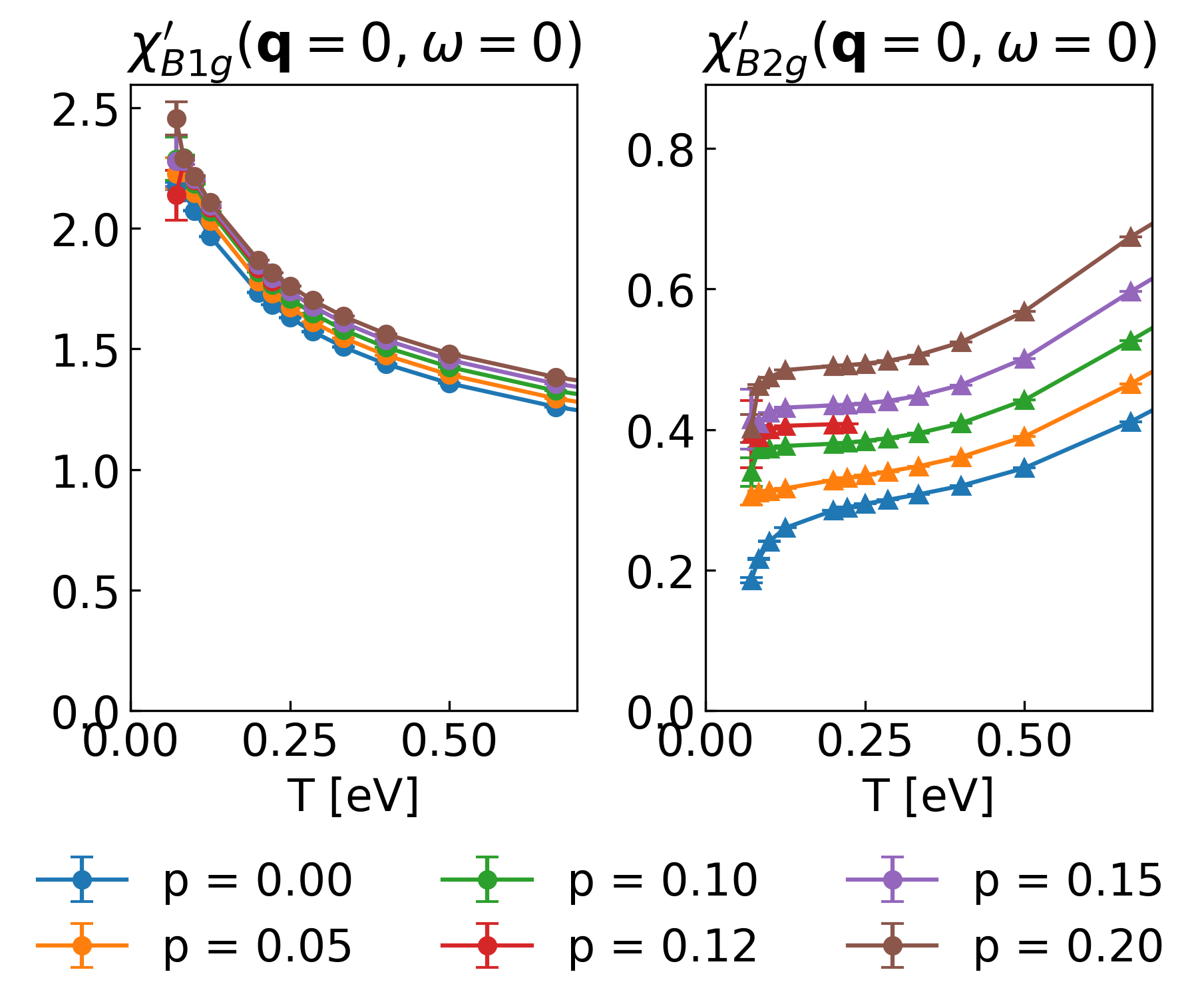}
    \caption{\textbf{Temperature evolution of bond-charge nematic susceptibilities from 8$\times$8 clusters} (a) Nearest neighbor Cu-O $B_{1g}$ susceptibility (b) Nearest neighbor O-O $B2g$ susceptibility for the $U_{dd} = 6$ eV parameter set. (a) Anisotropy in kinetic energy induced by a small perturbation in $t_{pd}$ in $x$ and $y$ direction increases with decreasing temperature. $\chi_{B_{1g}}$ grows faster below $T\simeq 0.4$~eV. (b) Anisotropy in kinetic energy induced by a small perturbation in $t_{pp}$ decreases with decreasing temperature. $\chi_{B_{2g}}$ decreases with temperature faster above the $T\sim0.4$ eV, which is likely due to the charge transfer energy, and further decreases below $T\sim 0.1$ eV, echoing the temperature evolution of intertwined stripes in Fig.~\ref{fig:temperature-locking}.}
    \label{fig:rho-and-E}
\end{figure}

This crossover 
is further corroborated by the temperature evolution of the $B_{1g}$ and $B_{2g}$ bond-charge nematic susceptibilities. Figure~\ref{fig:rho-and-E} (a) and (b) display the static nematic susceptibilities $\chi_{B_{1g}}(\omega = 0)$ and $\chi_{B_{2g}}(\omega = 0)$ defined in Sec.~\ref{sec:method}. These quantities measure the linear response of the kinetic energy to a small anisotropy in the hopping parameters, but excluding the contribution from the change in hopping parameters to the first order. Panel (a) demonstrates that $\chi_{B_{1g}}(\omega =0)$ grows as temperature decreases, accelerating notably below $T\simeq 0.2$~eV, which roughly coincides with the emergence of intertwined stripe patterns in the correlation functions. Panel (b) reveals that $\chi_{B_{2g}}(\omega =0)$ generally decreases with decreasing temperature, exhibiting a precipitous drop near $T\simeq 0.4$~eV. This initial drop corresponds to a rapid, temperature-driven increase in charge transport diffusivity \cite{3-band-transport}. Below $T \simeq 0.1$~eV, $\chi_{B_{2g}}$ undergoes another sharp downturn, corresponding to the onset of the diamond-like spin and charge stripe patterns observed in Fig.~\ref{fig:temperature-locking}. Together, these behaviors suggest that a diverging nematic susceptibility in a single dominant symmetry channel is closely tied to the formation of unidirectional stripes.\\

 Figures~\ref{fig:temperature-locking} and~\ref{fig:rho-and-E} jointly reveal that the $d$-orbital spin correlation and oxygen charge correlation are already mutually coupled at around $T\simeq 0.2$~eV, whereas the spontaneous tendency to form unidirectional stripes along the $d$-$p$ bonds only emerges at lower temperatures ($T\sim 0.1$~eV) for the $U_{dd} = 6$~eV parameter set. For the other two parameter sets, the fermion sign problem precludes accessing this low-temperature regime, leaving the $B_{2g}$ mode still significant at the lowest accessible temperatures. Nevertheless, the temperature evolution of both the nematic susceptibility and spatial correlation functions remains qualitatively similar across all sets within their respective accessible temperature ranges. Comparing across the Emery model parameters, while the $U_{dd} = 8.5$~eV parameter sets does not yet exhibit a spontaneous $B_{1g}$ stripe tendency on the square lattice at accessible temperatures, it produces more intense charge modulations on the $16\times 4$ clusters where the $C_4$ symmetry is externally broken. Consequently, the intrinsic amplitude of the charge modulation and its tendency towards $B_{1g}$ symmetry breaking must be understood as distinct properties of the intertwined spin and charge stripe. \\

\begin{figure}
    \centering
    \includegraphics[width=1.0\linewidth]{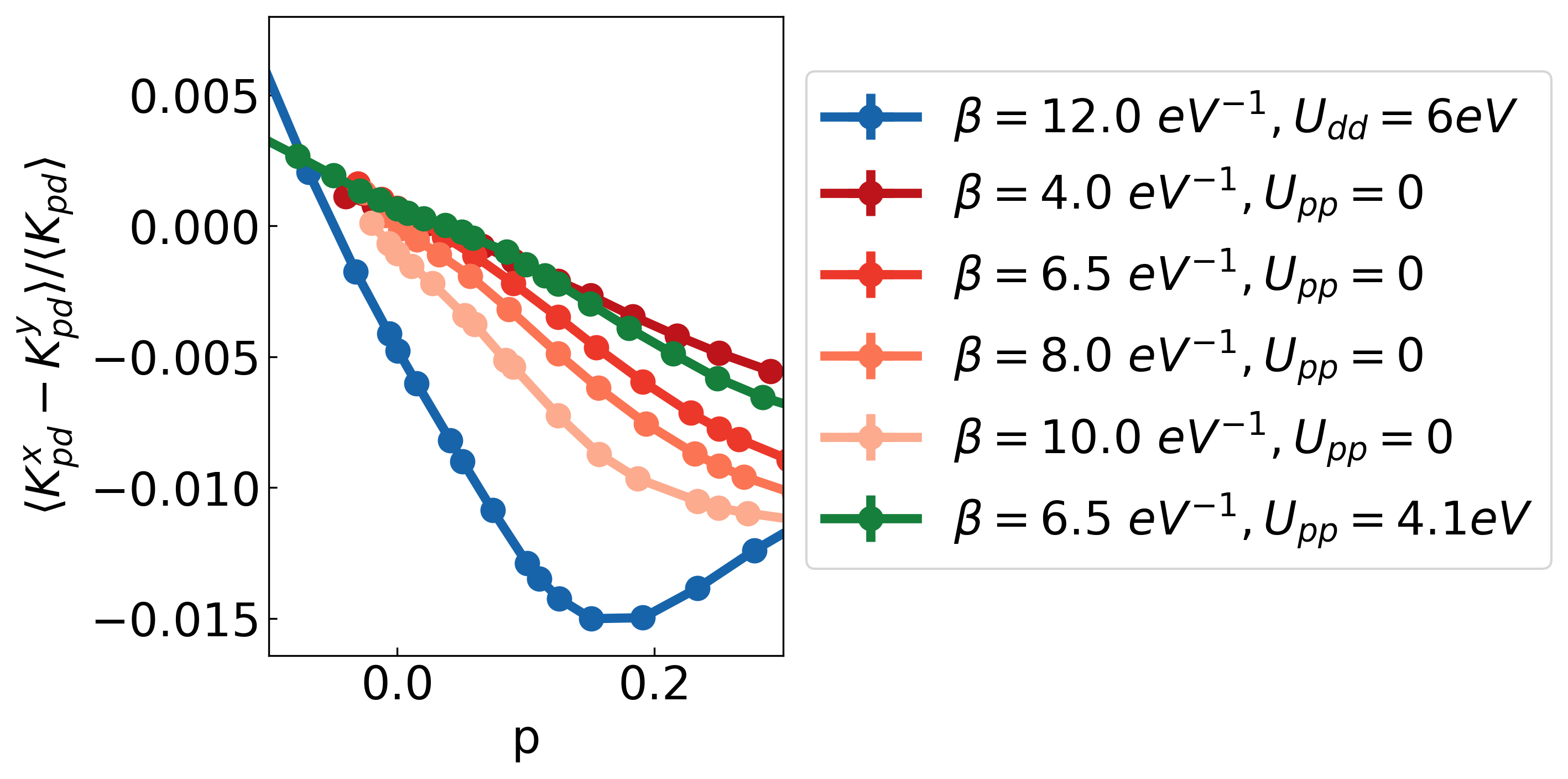}
    \caption{\textbf{Normalized kinetic energy anisotropy on $16 \times 4$ clusters at different hole doping $p$ and inverse temperature $\beta$} Positive values means kinetic energy is more negative along the $x$-direction.}
    \label{fig:kinetic-anisotropy}
\end{figure}

 Having established that the bond-charge nematic susceptibilities -- the tendency toward spontaneous kinetic energy anisotropy -- is linked to the formation of charge stripes, we now focus on the kinetic energy anisotropy in $16 \times 4$ clusters, where a unidirectional stripe pattern is visible from the static correlation function. As shown in Fig.~\ref{fig:kinetic-anisotropy}, the normalized kinetic energy anisotropies are negative for hole dopings $p > 0$. This indicates that the kinetic energy is enhanced along the $y$-direction, the direction of the antiferromagnetic domain wall, with the anisotropy growing in magnitude as temperature decreases. This finite-temperature observation aligns with ground-state DMRG results \cite{White_2015}, which demonstrates that vertical hopping amplitudes are larger than the horizontal ones near the antiferromagnetic domain walls. Notably, the doping dependence of this anisotropy is non-monotonic. For $U_{dd} = 6$~eV at $\beta = 12$~eV$^{-1}$, the kinetic energy anisotropy shows a pronounced minimum near $p = 0.15$. A similar low-temperature curvature also begins to develop for the $U_{dd} = 8.5$~eV, $U_{pp} = 0$ parameter set. While a comparable non-monotonic doping dependence for nematic susceptibility or stripe tendencies has been observed in the single-band Hubbard model and cuprates \cite{HubbardNematic, HubbardChargeStripe}, a direct connection remains yet to be established between nematic susceptibility and charge stripes. In contrast, our Emery model results demonstrate that the bond-charge nematicity and kinetic energy anisotropy are coupled with the formation of charge stripes.\\

 The $p$-orbital preference of the charge modulation and the kinetic energy asymmetry may help elucidate the microscopic origin of the charge stripes. From the perspective of perturbation theory, in the limit of $U_{dd}\to\infty$, one extra hole on a $p$ orbital in a half-filled Emery model induces a {\it ferromagnetic} correlation between its two neighboring $d$ orbitals. This interaction is due to the effective $d$-$p$ antiferromagnetic exchange given by $t_{pd}^2/(U_{pp} + \Delta_{pd})$ \cite{Lau2011}, which locally disrupts the antiferromagnetic background. Depending on whether the doped holes primarily occupy the $p_x$ or $p_y$ orbitals, two heuristic scenarios can be constructed. Consider a $16 \times4$ cluster where the antiferromagnetic domain wall is oriented along the $y$-direction. If the doped holes were to reside on the $p_x$ orbitals, the local $d$-$d$ and $d$-$p$ exchange interactions could be largely satisfied, and hole motion along the $x$-direction would fluctuate the antiferromagnetic domain wall. However, our numerical results for the Emery model turn out to be different: $p_y$ orbitals exhibit a higher hole occupation near the antiferromagnetic domain wall than $p_x$ orbitals, and the kinetic energy along the $y$ direction is more negative than along the $x$ direction, which frustrates the antiferromagnetic correlation along the $y$-axis. This preference of $p_y$ occupation indicates that the system gains more kinetic energy through the delocalization along the $y$-direction than the transverse fluctuations of the AFM domain walls.\\

\section{Conclusions}

 In summary, we report the 
real-space observation of intertwined spin and charge stripes at finite temperatures using DQMC on rectangular clusters, finding that they approximately satisfy the Yamada relation. This is achieved through exploring the balance between the fermion sign problem and the intrinsic tendency to form stripes with a few different cuprate-relevant parameter sets. Our ground-state DMRG simulation on $4$-leg ladders robustly corroborates these findings; both methods consistently show that the copper $d$-orbital spin stripes intertwine with the oxygen charge stripes, and that the charge modulation resides primarily on the $p$ orbitals oriented parallel to the rivers of charge.

We found that at finite temperatures, particularly for $U_{dd} = 8.5$~eV, a local perturbation to the charge transfer energy induces a charge modulation wavevector perpendicular to the corresponding $d$-$p$ bond, indicating that the charge modulation could be short-ranged and sensitive to impurities. Notably, it could provide a compelling microscopic mechanism for the short-ranged, randomly oriented stripe patches that have been recently observed in STM experiments \cite{Ye2023a}. Therefore, an interesting problem for future research would be to model the $C_4$ symmetry-breaking impurities and determine their effect on the charge and spin stripes.\\

By tracking the temperature evolution of the $C_4$ symmetric cluster, we uncovered evidence that spin-charge-locking in correlation functions emerges at temperatures significantly higher than the dominance of the $B_{1g}$ symmetry-breaking tendency. As the system cools, the tendency to develop kinetic energy anisotropy evolves concomitantly with the spin and charge correlation functions. Crucially, when the $C_4$ symmetry is externally broken, the kinetic energy is enhanced along the antiferromagnetic domain walls. Because the kinetic energy dictates the sum rule for the optical conductivity, this enhancement supports the notion of ``rivers of conducting charge" from experimental observations, where intertwined stripes play a key role in charge transport. \\ 

 Ultimately, establishing this detailed, orbital-resolved picture of fluctuating stripes at finite temperatures bridges a critical gap between ground-state numerical techniques and experimental observations. By demonstrating that the three-band Emery model intrinsically captures these complex, intertwined nematic and stripe phases—and by explicitly connecting the orbital-selective charge modulations to directional transport—our results provide a vital microscopic foundation. Unraveling these intermediate-temperature, symmetry-breaking tendencies is a necessary step toward comprehensively understanding the anomalous normal-state transport, the strange metal regime, and the eventual emergence of superconductivity in strongly correlated hole-doped cuprates.

\section{acknowledgments}
 The authors would like to acknowledge helpful conversations with Peizhi Mai and Steven Johnston. The work at Stanford and SLAC was supported by the US Department of Energy, Office of Basic Energy Sciences, Materials Sciences and Engineering Division, under Contract No. DE-AC02-76SF00515. 
Computational work was performed on the Sherlock cluster at Stanford University and on resources of the National Energy Research Scientific Computing Center, supported by the U.S. Department of Energy under contract DE-AC02-05CH11231.

\section{data availability}
The DQMC simulations are based on an open-source package at ~\cite{github}.

\bibliography{bib}

\clearpage
\appendix
\begin{widetext}

\renewcommand{\thefigure}{S\arabic{figure}} 
\setcounter{figure}{0}
\
\section{Methods}
\subsection{Density matrix renormalization group}

We have implemented a version of DMRG \cite{White1992,McCulloch2002} that respects the $SU(2) \equiv SU(2)_{spin}\otimes U(1)_{charge}$ symmetry of the model, i.e., full spin rotational symmetry and charge conservation. We keep bond dimensions of $SU(2)$ multiplets up to $D$=16,000 (equivalent to $m \approx 47,000$ $U(1)$ states). This ensures accurate results with a typical truncation error $\epsilon\approx 3\times 10^{-7}$. The calculations are on a $24 \times4$ cylinder for $1/6$ and $1/12$ hole doping, and $32\times4$ cylinder for $1/8$ hole doping. The doping labeled in the main text is the hole doping counted within the $16$ rungs from the center, which is close to the total hole doping. The charge and spin correlation functions presented are extrapolated to infinite bond dimension.

\subsection{Bond charge nematic susceptibilities}
 In the gauge where the Emery model is manifestly $C_4$ symmetric, in momentum space, the bond charge nematic susceptibilities explicitly show the form factors for $B_{1g}$ and $B_{2g}$ symmetry. Using the gauge transformation as explained in Ref.~\cite{3-band-transport}, the bond charge nematic susceptibilities are given by
\begin{align*}
    \begin{split}
    {\rho}_{B_{1g}} =& \sum_i \xi^{a}_i - \xi^{b}_i + \xi^{c}_i - \xi^{d}_{i}
    = \sum_{\mathbf{k},\sigma}[2\cos(k_{x}/2)(\widetilde{p}_{\mathbf{k},\sigma}^{x\dagger}\widetilde{d}_{\mathbf{k},\sigma} + \widetilde{d}_{\mathbf{k},\sigma}^{\dagger}\widetilde{p}_{\mathbf{k},\sigma}^{x})
    - 2\cos(k_{y}/2)(\widetilde{p}_{\mathbf{k},\sigma}^{y\dagger}\widetilde{d}_{\mathbf{k},\sigma} + \widetilde{d}_{\mathbf{k},\sigma}^{\dagger}\widetilde{p}_{\mathbf{k},\sigma}^{y})]
    \end{split}
\end{align*}
\begin{align*}
\begin{split}
    \rho_{B_{2g}} &= \xi^e_i - \xi^{f}_i + \xi^{g}_i - \xi^{h}_i
    = \sum_{\mathbf{k}}4\sin(k_x/2)\sin(k_y/2) (\tilde{p}_{\mathbf{k},\sigma}^{x\dagger}\tilde{p}_{\mathbf{k},\sigma}^{y} + \tilde{p}_{\mathbf{k},\sigma}^{y\dagger}\tilde{p}_{\mathbf{k},\sigma}^{x}) 
\end{split}
\end{align*}\\
\subsection{Determinant quantum Monte Carlo}
DQMC is numerically exact but limited by the fermion sign problem. The fermion sign of the Emery model depends on the parameters, and gets smaller when $U_{dd}$, $U_{pp}$, $\Delta_{pd}$ increase or the temperature decreases. Figure~\ref{fig:sign} shows the average sign.\\

\begin{figure}
    \centering
    \includegraphics[width=0.9\linewidth]{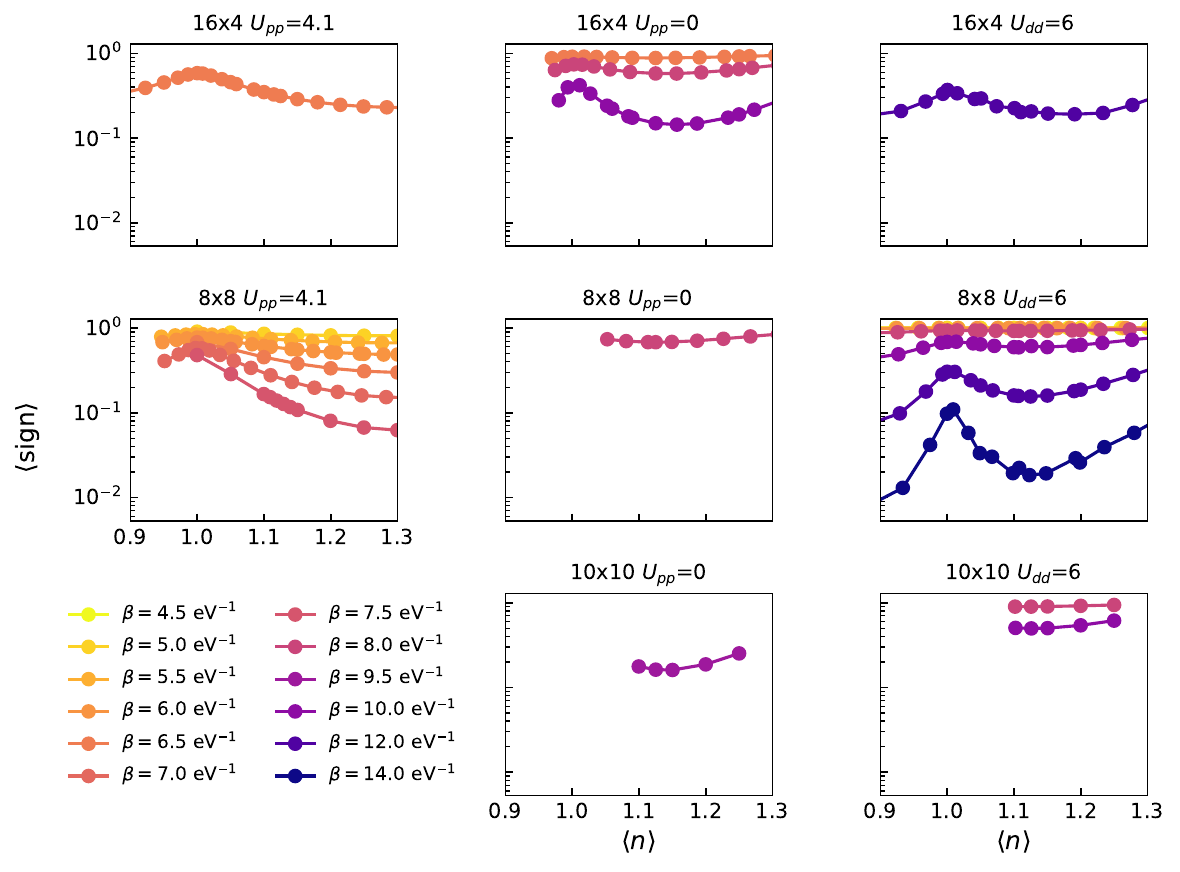}
    \caption{Averaged fermion sign for various lattice geometries at different temperatures and dopings}
    \label{fig:sign}
\end{figure}
Due to the exponentially decaying charge correlation signal and the slow convergence, we do not explore a regime with even smaller average sign. The number of measurement sweeps and Markov chains is shown in Table~\ref{tab:measurements}. Doping levels are tuned by adjusting the chemical potential to the accuracy of $\sim 10^{-4}$ in the average density. \\

\begin{table}[h]
\centering
\begin{tabular}{ccccccc}
\toprule
Calc.\ type & Size & $\beta$ & $n$ & \# chains & uneqlt. meas./chain & sweeps/ uneqlt. meas. \\
\midrule
$U_{pp}$ = 4.1 eV & $16\times 4$ & 6.5 & 1.050 & 395 & 75,000 & 8 \\
$U_{pp}$ = 4.1 eV & $16\times 4$ & 6.5 & 1.100, 1.125, 1.150 & 990--992 & 75,000 & 8 \\
\midrule
$U_{pp}$ = 0 eV & $16\times 4$ & 8.0 & 1.125 & 1,989 & 37,500 & 16 \\
$U_{pp}$ = 0 eV & $16\times 4$ & 8.0 & 1.250 & 1,000 & 37,500 & 16 \\
\midrule
$U_{pp}$ = 0 eV & $16\times 4$ & 10.0 & 1.060, 1.125, 1.190, 1.250 & 906--998 & 45,000 & 16 \\
$U_{pp}$ = 0 eV & $16\times 4$ & 10.0 & 1.090 & 390 & 75,000 & 8 \\
\midrule
$U_{dd}$ = 6 eV & $16\times 4$ & 12.0 & 1.050 & 917 & 37,500 & 16 \\
$U_{dd}$ = 6 eV & $16\times 4$ & 12.0 & 1.100, 1.125, 1.150 & 1,991--2,000 & 37,500 & 16 \\
$U_{dd}$ = 6 eV & $16\times 4$ & 12.0 & 1.190 & 1,000 & 37,500 & 16 \\
\midrule
$U_{pp}$ = 4.1 eV & $8\times 8$ & 5.0 & 1.050, 1.100, 1.150, 1.200, 1.250, 1.300 & 40 & 37,500 & 16 \\
\midrule
$U_{pp}$ = 4.1 eV & $8\times 8$ & 6.0 & 1.050, 1.100, 1.150, 1.200, 1.250, 1.300 & 372--380 & 37,500 & 16 \\
\midrule
$U_{pp}$ = 4.1 eV & $8\times 8$ & 6.5 & 1.050, 1.100, 1.150, 1.200 & 267--390 & 37,500 & 16 \\
$U_{pp}$ = 4.1 eV & $8\times 8$ & 6.5 & 1.250, 1.300 & 465--472 & 37,500 & 16 \\
\midrule
$U_{pp}$ = 4.1 eV & $8\times 8$ & 7.5 & 1.050, 1.100, 1.150, 1.200, 1.250, 1.300 & 497--500 & 37,500 & 16 \\
\midrule
$U_{dd}$ = 6 eV & $8\times 8$ & 4.5 & 1.050, 1.100, 1.125, 1.150, 1.200 & 39--40 & 37,500 & 16 \\
\midrule
$U_{dd}$ = 6 eV & $8\times 8$ & 5.0 & 1.050, 1.100, 1.125, 1.150, 1.200 & 88--136 & 37,500 & 16 \\
\midrule
$U_{dd}$ = 6 eV & $8\times 8$ & 6.0 & 1.125 & 100 & 37,500 & 16 \\
\midrule
$U_{dd}$ = 6 eV & $8\times 8$ & 8.0 & 1.050, 1.100 & 378--381 & 37,500 & 16 \\
$U_{dd}$ = 6 eV & $8\times 8$ & 8.0 & 1.125 & 188 & 37,500 & 16 \\
$U_{dd}$ = 6 eV & $8\times 8$ & 8.0 & 1.150, 1.200 & 379--388 & 37,500 & 16 \\
\midrule
$U_{dd}$ = 6 eV & $8\times 8$ & 10.0 & 1.050, 1.100, 1.125, 1.150, 1.200 & 191--268 & 37,500 & 16 \\
\midrule
$U_{dd}$ = 6 eV & $8\times 8$ & 12.0 & 1.050, 1.100, 1.125, 1.150, 1.200 & 476--491 & 37,500 & 16 \\
\midrule
$U_{dd}$ = 6 eV & $8\times 8$ & 14.0 & 1.050, 1.100, 1.125, 1.150, 1.200 & 680--684 & 37,500 & 16 \\
\bottomrule
\end{tabular}
\caption{\textbf{Simulation parameters of DQMC calculations} ``uneqlt. meas." stands for unequal time measurements. ``Chains" refers to independent Markov chains.}
\label{tab:measurements}
\end{table}

Jackknife resampling is used to estimate the standard errors. The spacing in the imaginary time grid $\delta\tau \leq 0.05~$eV$^{-1}$ for all simulations with $U_{dd} = 8.5$~eV and $\delta\tau\leq 0.08~$eV $^{-1}$ for some simulations with $U_{dd} = 6$~eV. 

\subsection{Monte Carlo sampling}
 In a system with translational symmetry, $\chi_{ij} = \chi_{i'j'}$ if $\mathbf{r_i} - \mathbf{r}_j = \mathbf{r}_{i'} - \mathbf{r}_{j'}$, where $i,i'$ and $j,j'$ belong to the same type of orbital. For the charge correlation function, although $\langle n_i(\tau)\rangle = \langle n_i(0)\rangle$, the Monte Carlo samples for different time slices are generally different. The correlation function for each time slice are evaluated by $\langle n_i(\tau)n_j(0)\rangle  - \langle n_i(\tau)\rangle \langle n_j(0)\rangle$ from the Monte Carlo data, which reduces the number of samples required compared to other estimators of the orbital resolved density. In addition, we utilized all spatial symmetries and the symmetry given by the cyclic property of the trace $\langle n_i(\tau)n_j(0) \rangle = \langle n_j(\beta - \tau)  n_i(0)\rangle$, which further reduces the statistical error.\\

\subsection{Symmetry of correlation functions}
 Due to the $C_2$ point group symmetry of the oxygen orbitals in the Emery model, the $p_x$-$p_x$ and $p_y$-$p_y$ components of the charge correlation functions have $C_2$ symmetry, where the $C_4$ symmetry is explicitly broken by the perturbation of the charge transfer energy on $p_x$ or $p_y$ orbitals, but they are related by 90 degree rotation to each other. On a $C_4$ symmetric square lattice, we can rotate our coordinate system and exchange the $p_x$ and $p_y$ orbital labels, {\it i.e.} $\chi_{c, ij}^{xy}(\omega = 0) = \chi_{c,\pi(i)\pi(j)}^{yx}(\omega = 0)$, where $\pi(\cdot)$ maps the orbital location in the original coordinate system to a new coordinate system with a $90^\circ$ rotation. The cyclic property of trace implies that we have $\chi_{c,ij}^{xy}(\omega = 0) = \chi_{c,ji}^{yx}(\omega = 0)$. Comparing the two equalities above, we know that $\chi_{c,ij}^{xy}$ is symmetric under $90^\circ$ rotation, and hence $C_4$ symmetric.

\subsection{Finite size effect}
 We simulate a few dopings near the lowest accessible temperatures, limited by the Fermion sign problem for $10\times 10$ clusters with $U_{dd} = 6$~eV and $U_{dd} = 8.5$~eV ($U_{pp} = 0$). The staggered static spin correlation functions and oxygen charge correlation functions are shown in Fig.~\ref{fig:10x10} and \ref{fig:10x10-Upp0}. The $\chi_c^{yy}$ is qualitatively similar to the $8\times 8$ clusters at the lowest accessible temperatures, which show positive charge correlation regions above and below the reference point, but even clearer. $\chi_c^{xy}$ for $U_{dd} = 6$~eV shows the $B_{1g}$ and $B_{2g}$ patterns, while the signal for $U_{dd} = 8.5$~eV, $U_{pp} = 0$ is mostly below the large error bar, but is not in contradiction to the $8\times 8$ data.

\begin{figure*}
    \centering
    \includegraphics[width=1.0\linewidth]{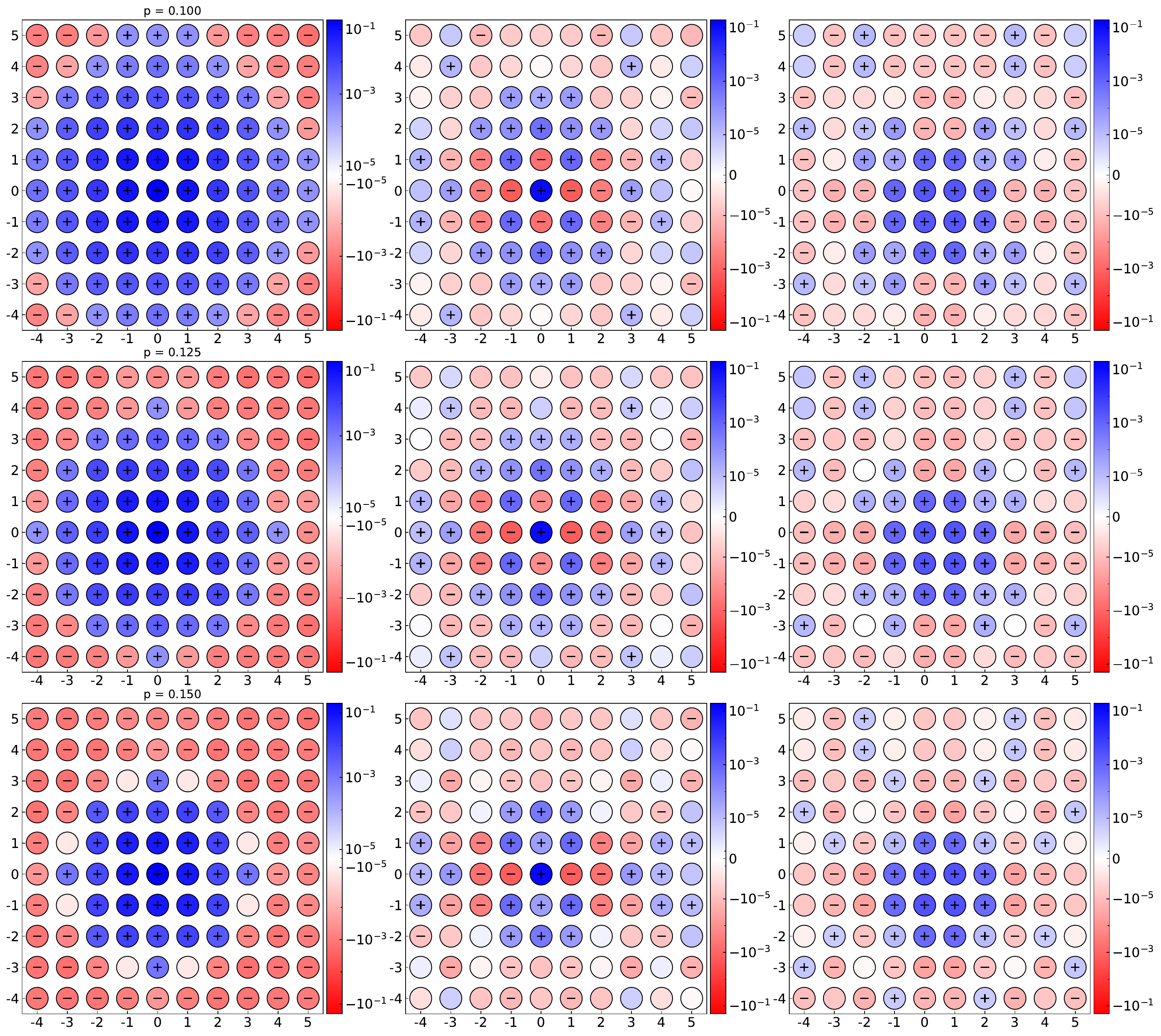}
    \caption{\textbf{Staggered spin correlation (first column) function and oxygen charge correlation functions (second column: $\chi_{c,ij}^{yy}$, third column: $\chi_{c,ij}^{xy}$) on $10\times10$ cluster} for the $U_{dd} = 6$~eV parameter set at $\beta = 10$~eV$^{-1}$. }
    \label{fig:10x10}
\end{figure*}

\begin{figure*}
    \centering
    \includegraphics[width=1.0\linewidth]{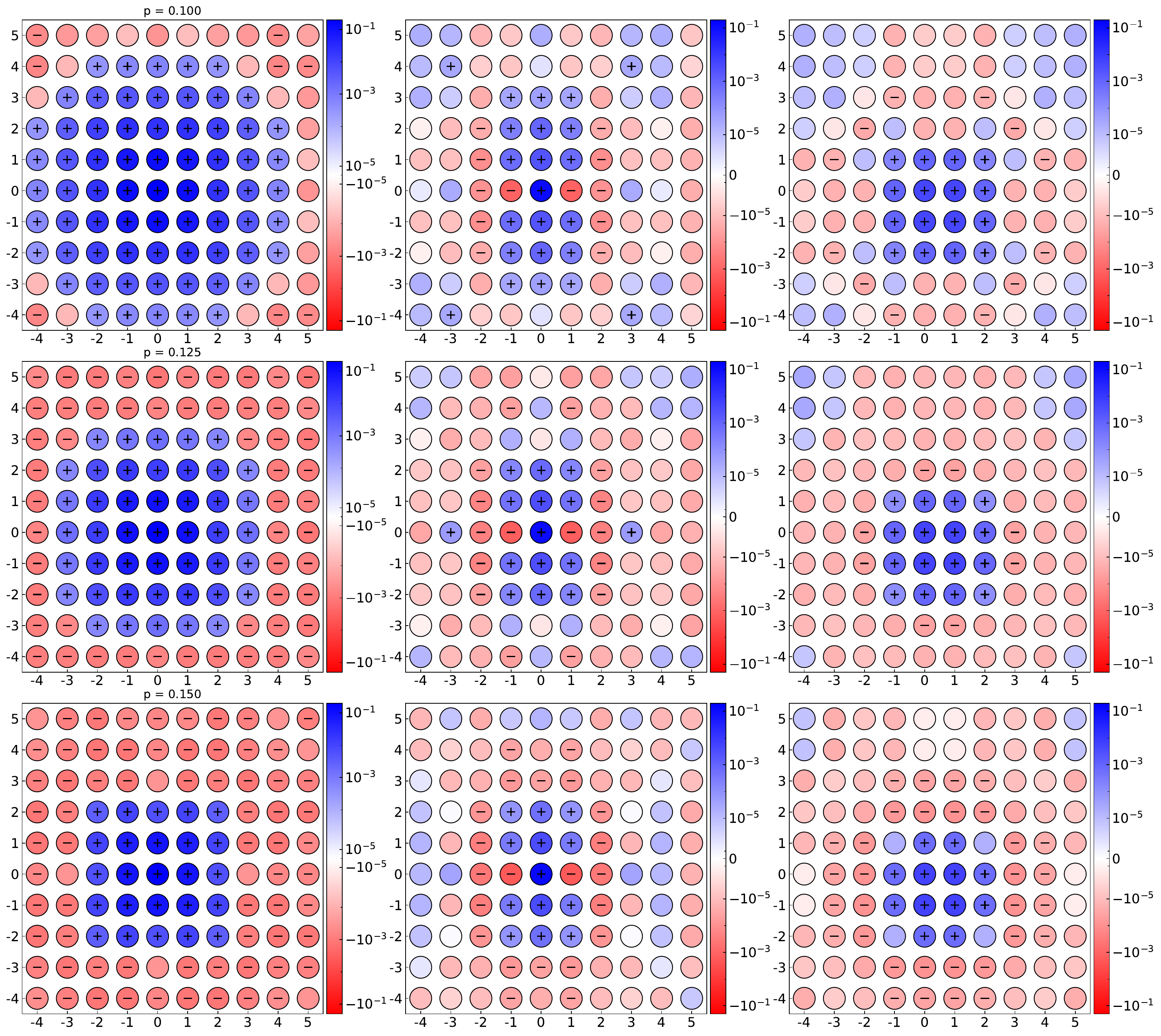}
    \caption{\textbf{Staggered spin correlation (first column) function and oxygen charge correlation functions (second column: $\chi_{c,ij}^{yy}$, third column: $\chi_{c,ij}^{xy}$) on $10\times10$ cluster} for the $U_{dd} = 8.5$~eV, $U_{pp} = 0$ parameter set at $\beta = 9.5$~eV$^{-1}$. }
    \label{fig:10x10-Upp0}
\end{figure*}

\section{Supplemental data}
\subsection{Orbital dependence of charge correlation}
A few components of the static charge susceptibility $\chi_c(\mathbf{q} = 0,\omega = 0)$ are plotted for each set of the Emery model parameters in Fig.~\ref{fig:nnk},~\ref{fig:nnk-Upp0}, and \ref{fig:nnk-Upp41}. $\chi_c^{dd}(\mathbf{q}, \omega = 0)$ is qualitatively the same as Ref.~\cite{PeizhiThreeband} for all three parameter sets. $\chi_c^{yy}(\mathbf{q})$ generally have a more significant double peak structure than $\chi_c^{xx}(\mathbf{q})$, and the doping dependence of the modulation wavevector is more significant.\\

\subsection{Antiferromagnetic exchange for different model parameters}

\begin{figure}
    \centering
    \includegraphics[width=0.8\linewidth]{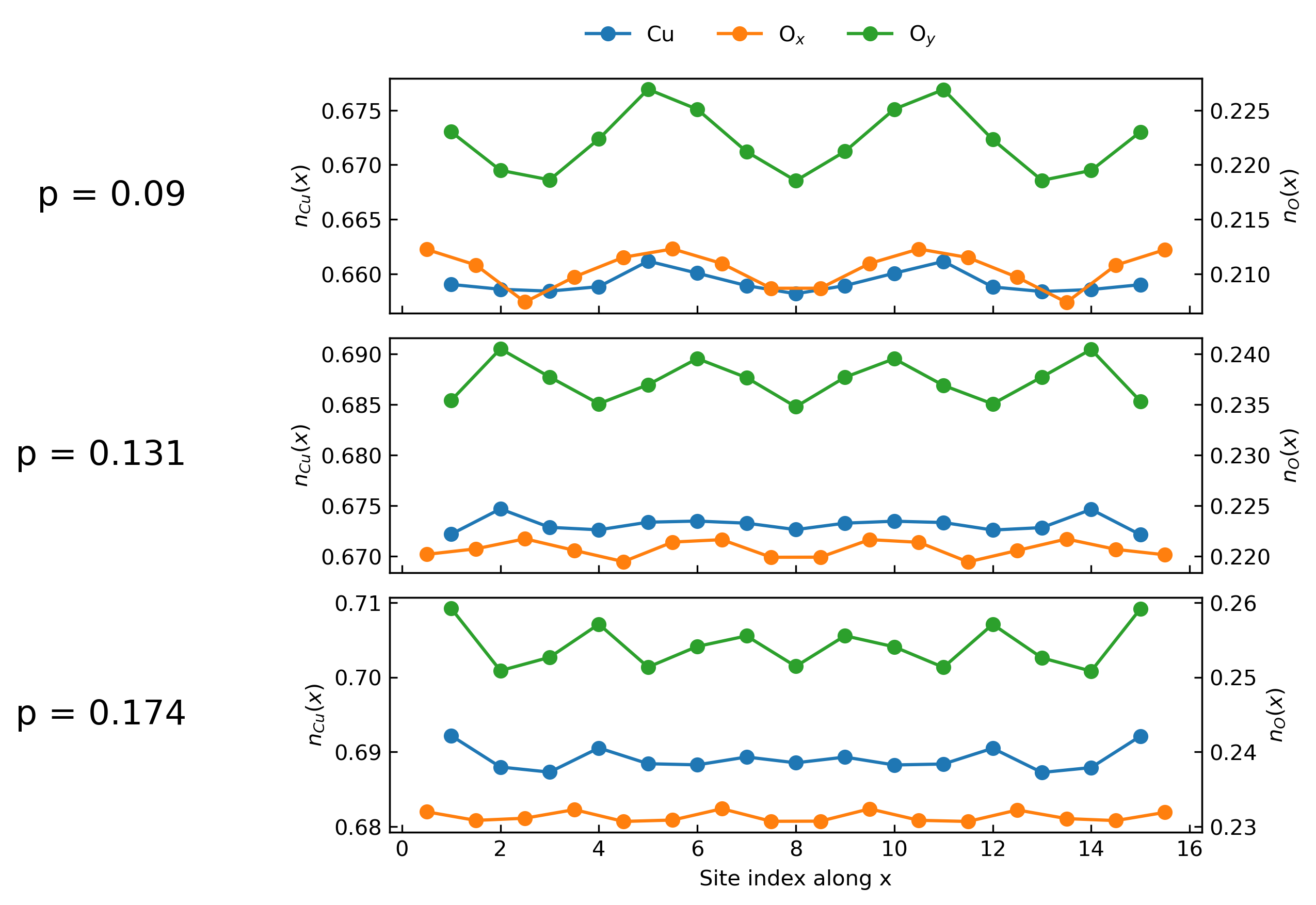}
    \caption{\textbf{DMRG densities within the inner 16 rungs of the cylinder for $U_{dd} = 6$ eV, $U_{pp} = 0$, $\Delta_{pd} = 3$ eV} \\ Figure S6 in both Refs.~\cite{threebandSpinStripe} and \cite{Ponsioen2023} show the density profile for the $U_{dd} = 4.1$~eV parameter set, where the maximum charge density variation  on $p_y$ orbital $\max_{i,j}| n_{y}(\mathbf{r}_{i}) - n_{y}(\mathbf{r}_{j})|$ is around 0.02. Here, $\max_{i,j}| n_{y}(\mathbf{r}_{i}) - n_{y}(\mathbf{r}_{j})|\simeq 0.01$, regardless of the actual hole doping.}
    \label{fig:DMRG}
\end{figure}

\begin{figure*}[hbt!]
    \centering
    \includegraphics[width=1.0\linewidth]{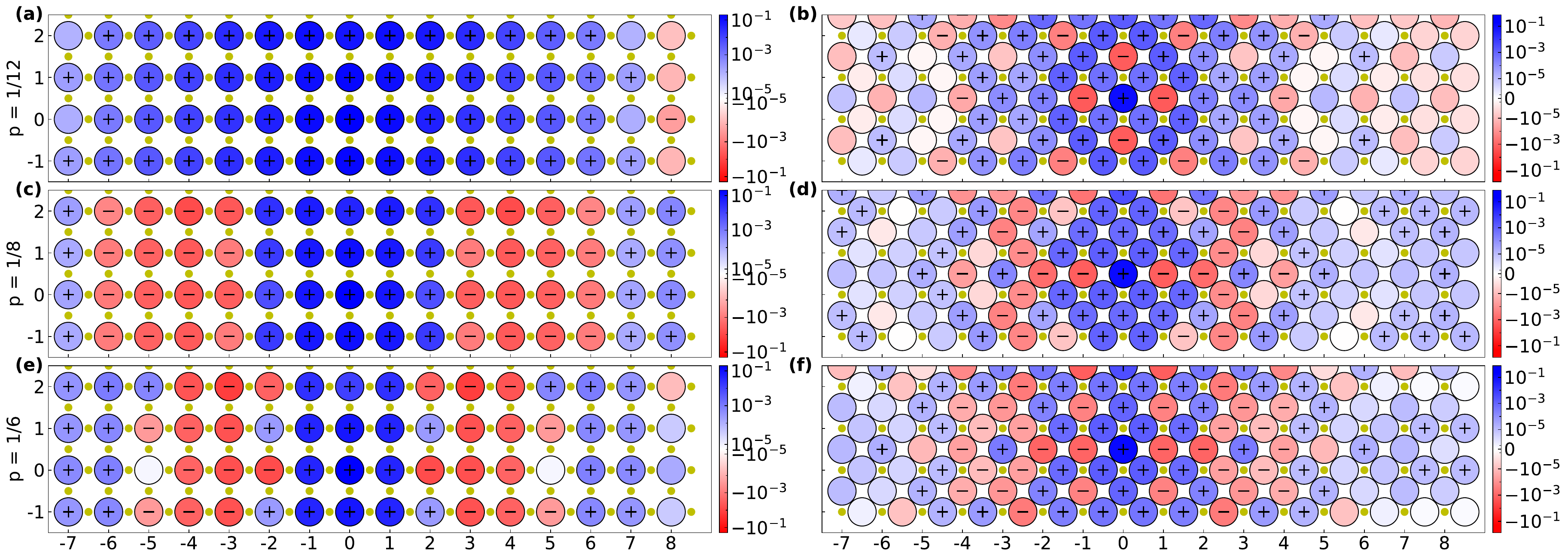}
    \caption{\textbf{Doping dependence of DQMC spin and charge stripes} \textbf{(a, c, e)}: staggered spin-spin correlation on copper $d$ orbitals for increasing hole dopings, in reference to the $d$ orbital at $(0,0)$. \textbf{(b, d, f)}: static connected charge-charge correlation on oxygen $p_{x}$ and $p_{y}$ orbitals in reference to the $p_y$ orbital at $(0, 1/2)$ for increasing dopings.  The value of the correlation functions is represented by the color in the large circles, and small yellow dots note the locations of other orbitals on the lattice. The charge correlation between the $p_y$ orbital at $(0, 1/2)$ and other $p$ ($d$) orbitals is represented by the color in the large (small) circles. The presence of plus (+) and minus (-) signs indicates that the data are more than $2$ standard errors from $0$. The static charge correlation pattern has a period that changes concomitantly with the antiferromagnetic domain walls in the spin-spin correlation. Despite the elevated temperature, DQMC shows spin and charge correlations that are consistent with the intertwined stripe in the DMRG ground state in Fig.~\ref{fig:dmrg}. Temperature and three-band parameters in this plot are $\beta = 12.0$ eV$^{-1}$, $U_{dd} = 6~e$V, $U_{pp} = 0$, $\Delta_{pd} = 3.0~e$V.}
    \label{fig:doping-spin-locking}
\end{figure*}

\begin{figure*}
    \includegraphics[width=1.0\linewidth]{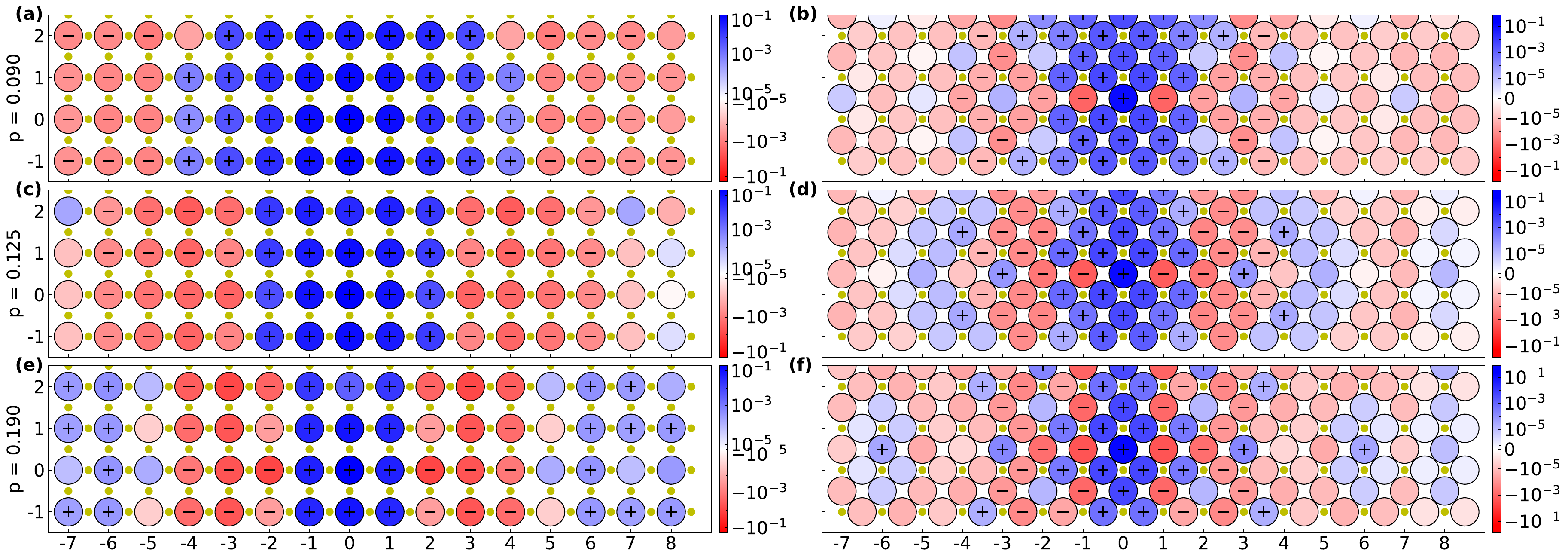}
    \caption{Same as Fig.~\ref{fig:doping-spin-locking} but for $U_{dd} = 8.5$~eV, $U_{pp} = 0$, at $\beta = 10.0$~eV$^{-1}$. The charge stripe pattern is clearer in real space compared to Fig.~\ref{fig:doping-spin-locking}.}
    \label{fig:doping-spin-locking-Upp0}
\end{figure*}

\begin{figure*}
    \includegraphics[width=1.0\linewidth]{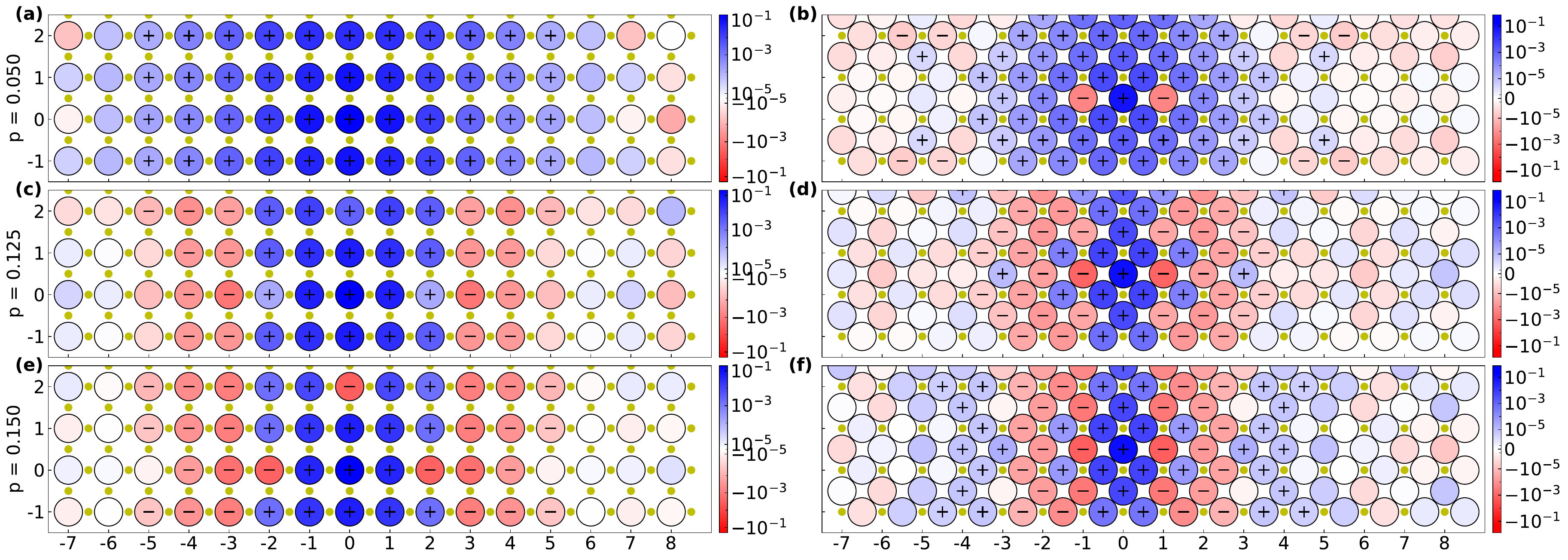}
    \caption{Same as Fig.~\ref{fig:doping-spin-locking} but for $U_{dd} = 8.5$~eV, $U_{pp} = 4.1$~eV, at $\beta = 6.5$~eV$^{-1}$. The charge stripe pattern is clearer in real space compared to Fig.~\ref{fig:doping-spin-locking}.}
    \label{fig:doping-spin-locking-Upp41}
\end{figure*}

\begin{figure*}
    \includegraphics[width=1.0\linewidth]{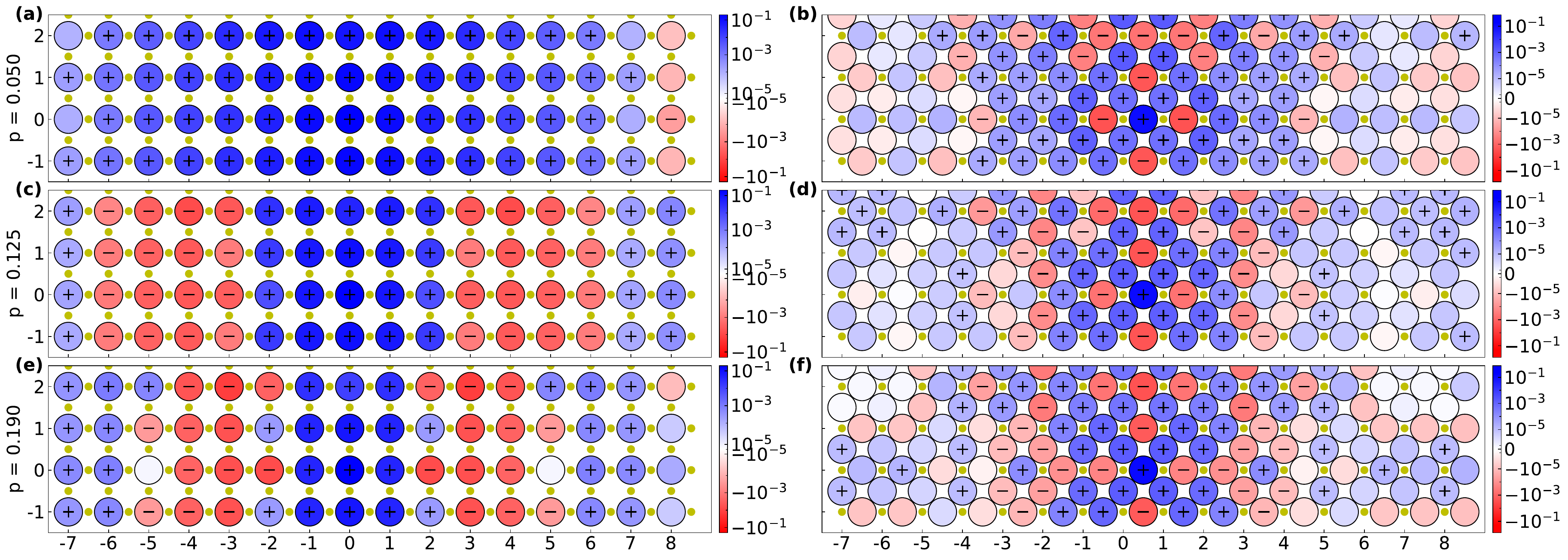}
    \caption{Same as Fig.~\ref{fig:doping-spin-locking} but in reference to $p_x$}
    \label{fig:doping-spin-locking-x}
\end{figure*}

\begin{figure*}
    \includegraphics[width=1.0\linewidth]{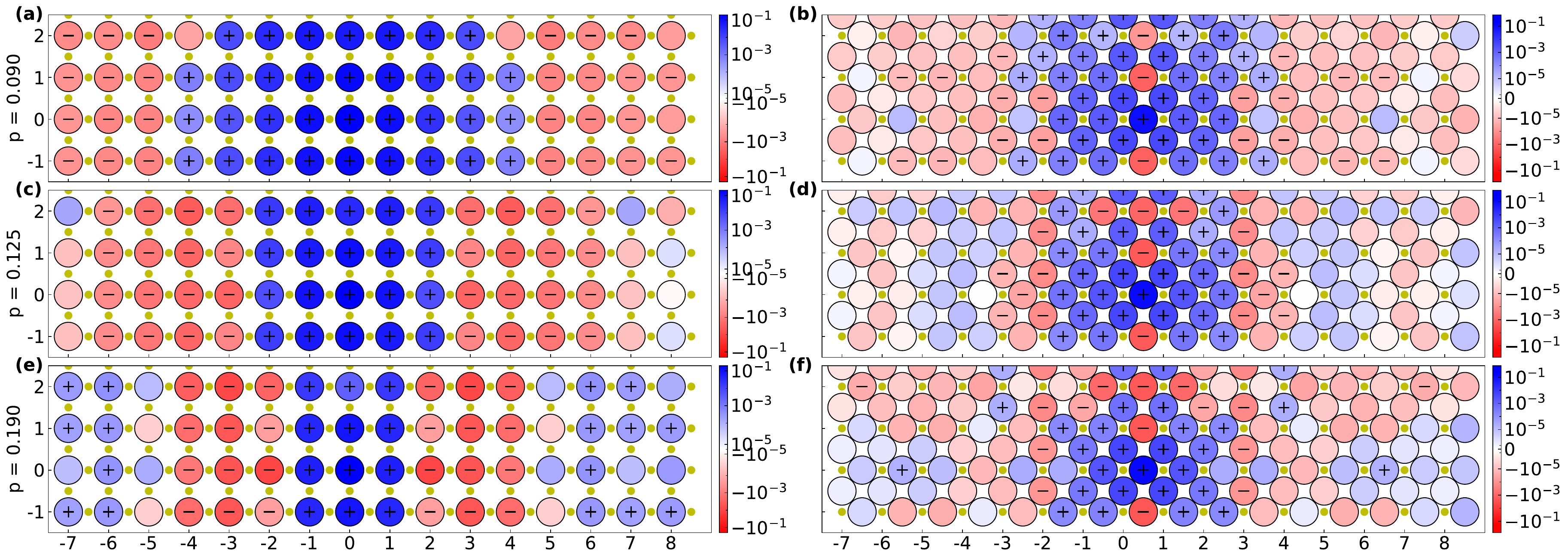}
    \caption{Same as Fig.~\ref{fig:doping-spin-locking-x} but for $U_{dd} = 8.5$~eV, $U_{pp} = 0$, at $\beta = 10.0$~eV$^{-1}$}
    \label{fig:doping-spin-locking-x-Upp0}
\end{figure*}

\begin{figure*}
    \includegraphics[width=1.0\linewidth]{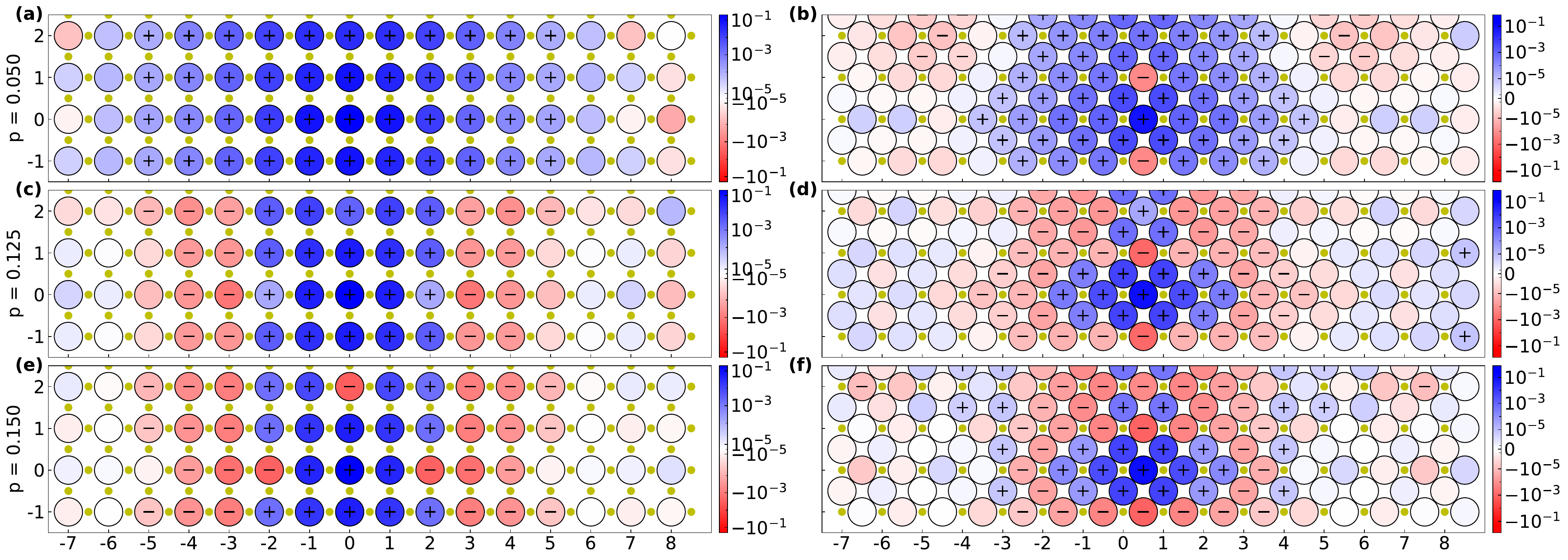}
    \caption{Same as Fig.~\ref{fig:doping-spin-locking-x} but for $U_{dd} = 8.5$~eV, $U_{pp} = 4.1$~eV, at $\beta = 6.5$~eV$^{-1}$}
    \label{fig:doping-spin-locking-x-Upp41}
\end{figure*}

\begin{figure*}
    \includegraphics[width=1.0\linewidth]{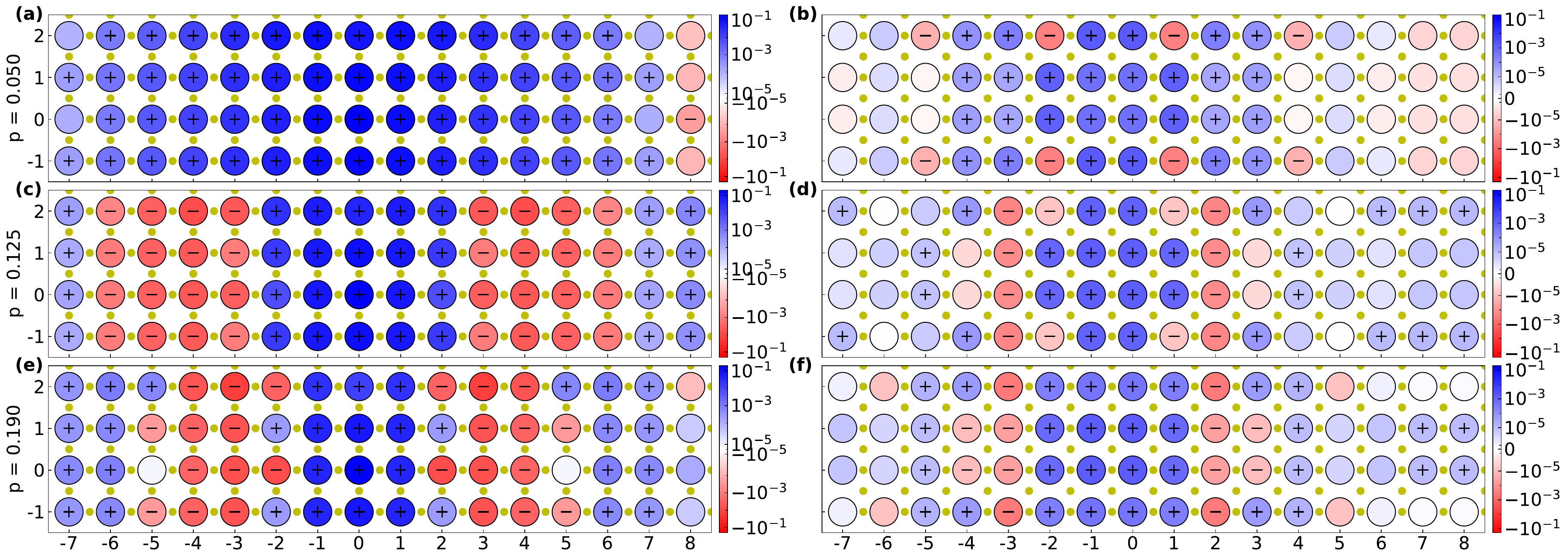}
    \caption{Same as Fig.~\ref{fig:doping-spin-locking} but shows charge density correlation between $p_x$ orbitals with respect to a $p_y$ orbital. The yellow dots stand for the positions of the $p_x$ and $d$ orbitals that are not shown. Stripe-like patterns are visible, although the doping dependence is not as clear.}
    \label{fig:doping-spin-locking-xy}
\end{figure*}

\begin{figure*}
    \includegraphics[width=1.0\linewidth]{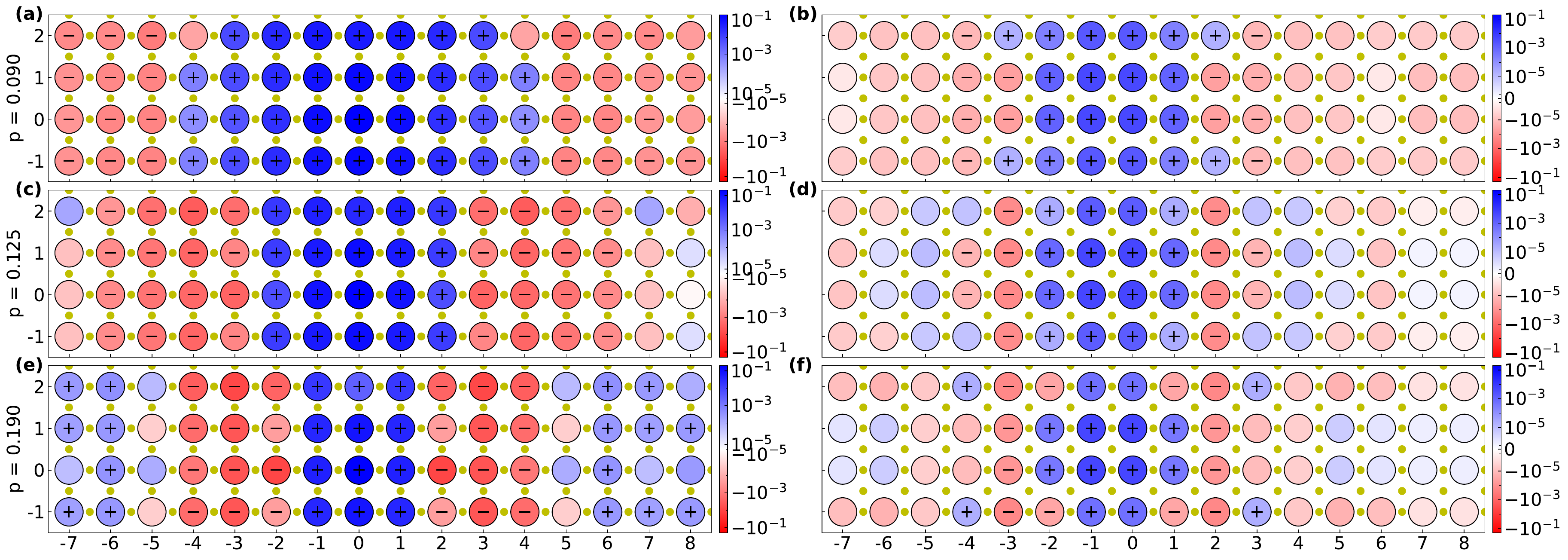}
    \caption{Same as Fig.~\ref{fig:doping-spin-locking-xy} but for $U_{dd} = 8.5$~eV, $U_{pp} = 0$, at $\beta = 10.0$~eV$^{-1}$. Charge correlation shows stripe-like patterns that move with the antiferromagnetic domain wall.}
    \label{fig:doping-spin-locking-xy-Upp0}
\end{figure*}

\begin{figure*}
    \includegraphics[width=1.0\linewidth]{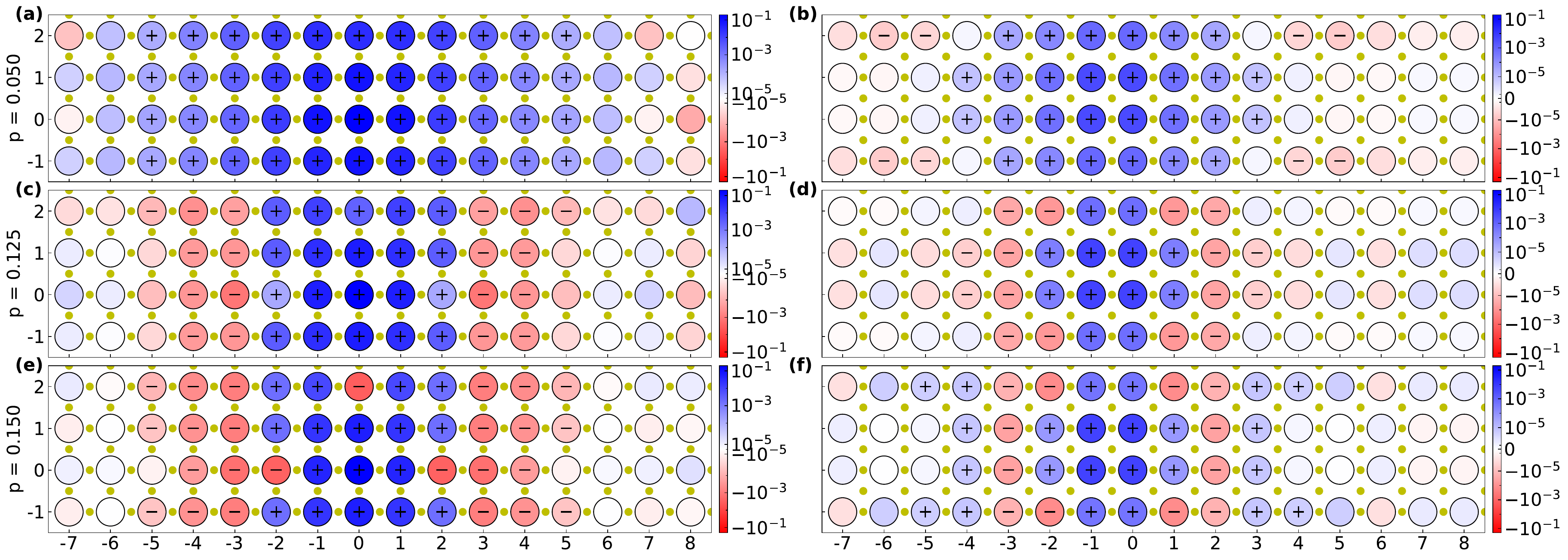}
    \caption{Same as Fig.~\ref{fig:doping-spin-locking-xy} but for $U_{dd} = 8.5$~eV, $U_{pp} = 4.1$~eV, at $\beta = 6.5$~eV$^{-1}$. Charge correlation shows stripe-like patterns that move with the antiferromagnetic domain wall.}
    \label{fig:doping-spin-locking-xy-Upp}
\end{figure*}

\begin{figure*}
    \includegraphics[width=1.0\linewidth]{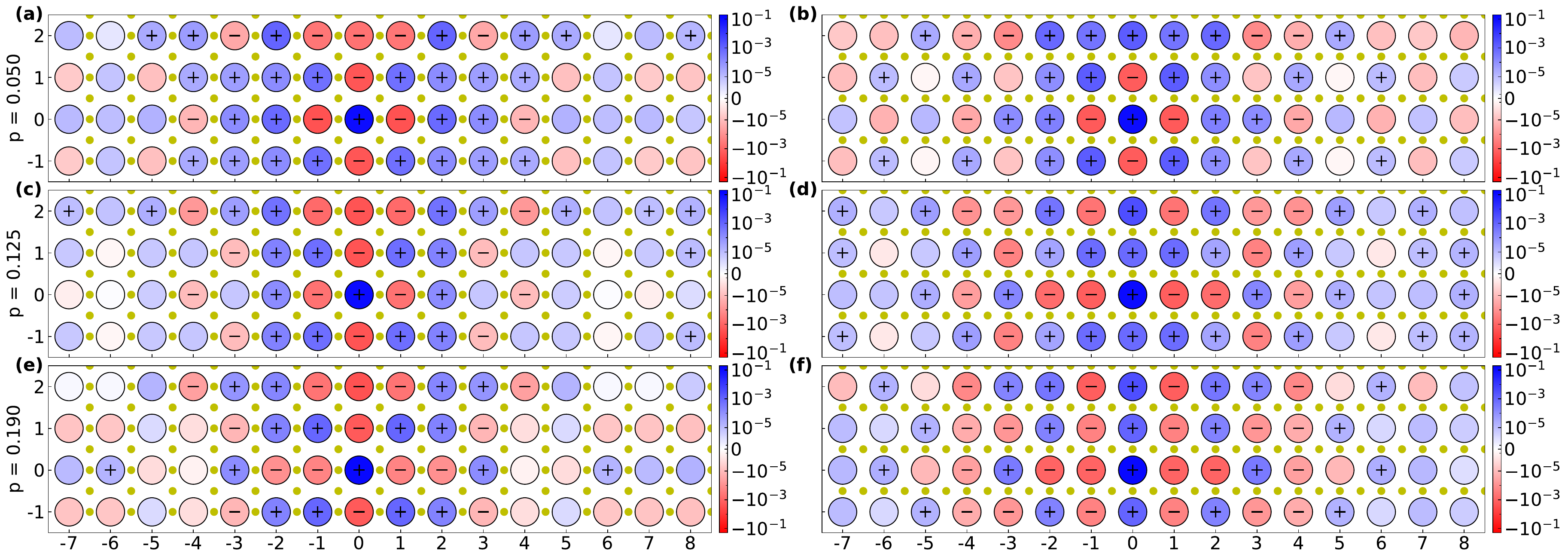}
    \caption{Same as Fig.~\ref{fig:doping-spin-locking} but shows charge density correlation between $p_x$ orbitals on the left and $p_y$ orbitals on the right. No clear pattern is visible on the $p_x$ orbitals. $p_y$ orbitals are not conclusive.}
    \label{fig:doping-spin-locking-yy}
\end{figure*}

\begin{figure*}
    \includegraphics[width=1.0\linewidth]{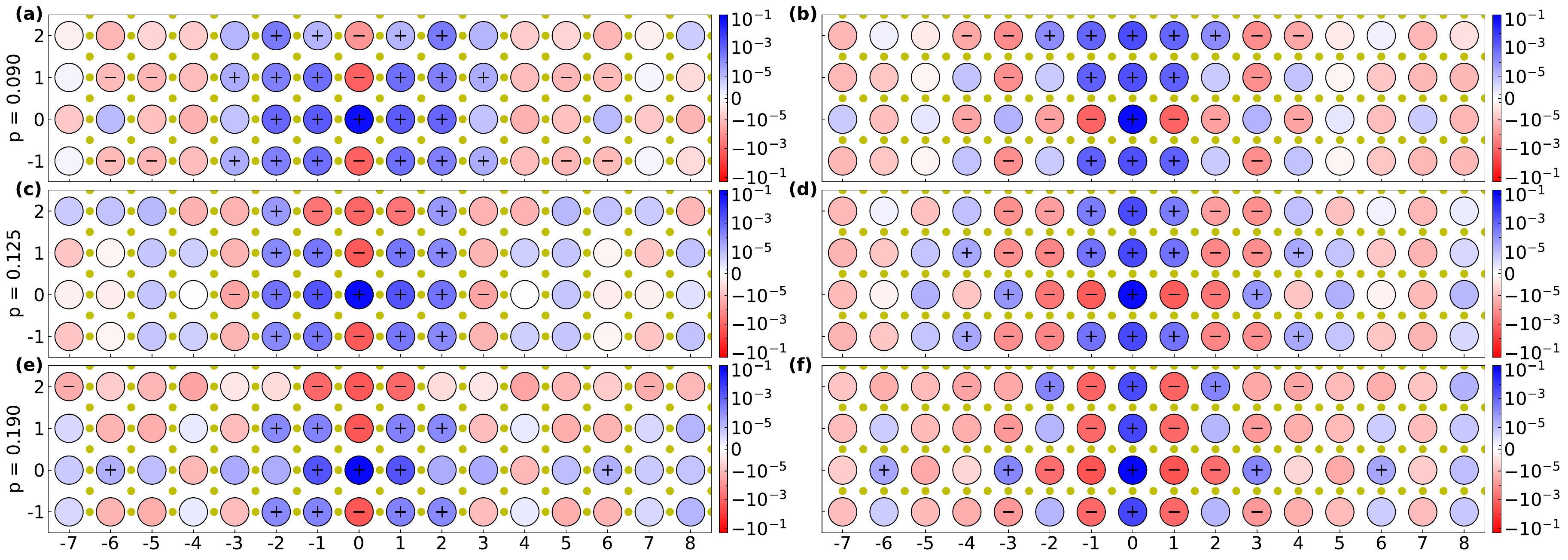}
    \caption{Same as Fig.~\ref{fig:doping-spin-locking-yy} but for $U_{dd} = 8.5$~eV, $U_{pp} = 0$, at $\beta = 10.0$~eV$^{-1}$. The stripe pattern is clearer compared to Fig.~\ref{fig:doping-spin-locking-yy}.}
    \label{fig:doping-spin-locking-yy-Upp0}
\end{figure*}

\begin{figure*}
    \includegraphics[width=1.0\linewidth]{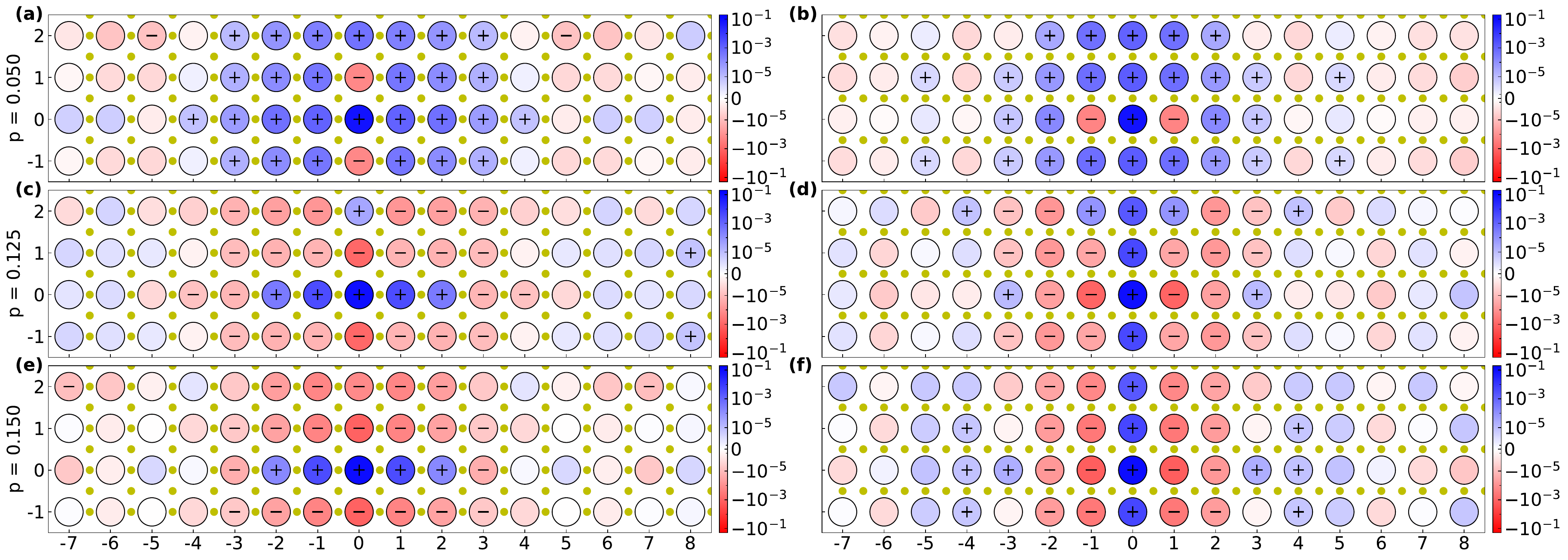}
    \caption{Same as Fig.~\ref{fig:doping-spin-locking-yy} but for $U_{dd} = 8.5$~eV, $U_{pp} = 4.1$~eV, at $\beta = 6.5$~eV$^{-1}$. The stripe pattern is clearer than Fig.~\ref{fig:doping-spin-locking-yy} and \ref{fig:doping-spin-locking-yy-Upp0}.}
    \label{fig:doping-spin-locking-yy-Upp}
\end{figure*}

\begin{figure*}
    \includegraphics[width=1.0\linewidth]{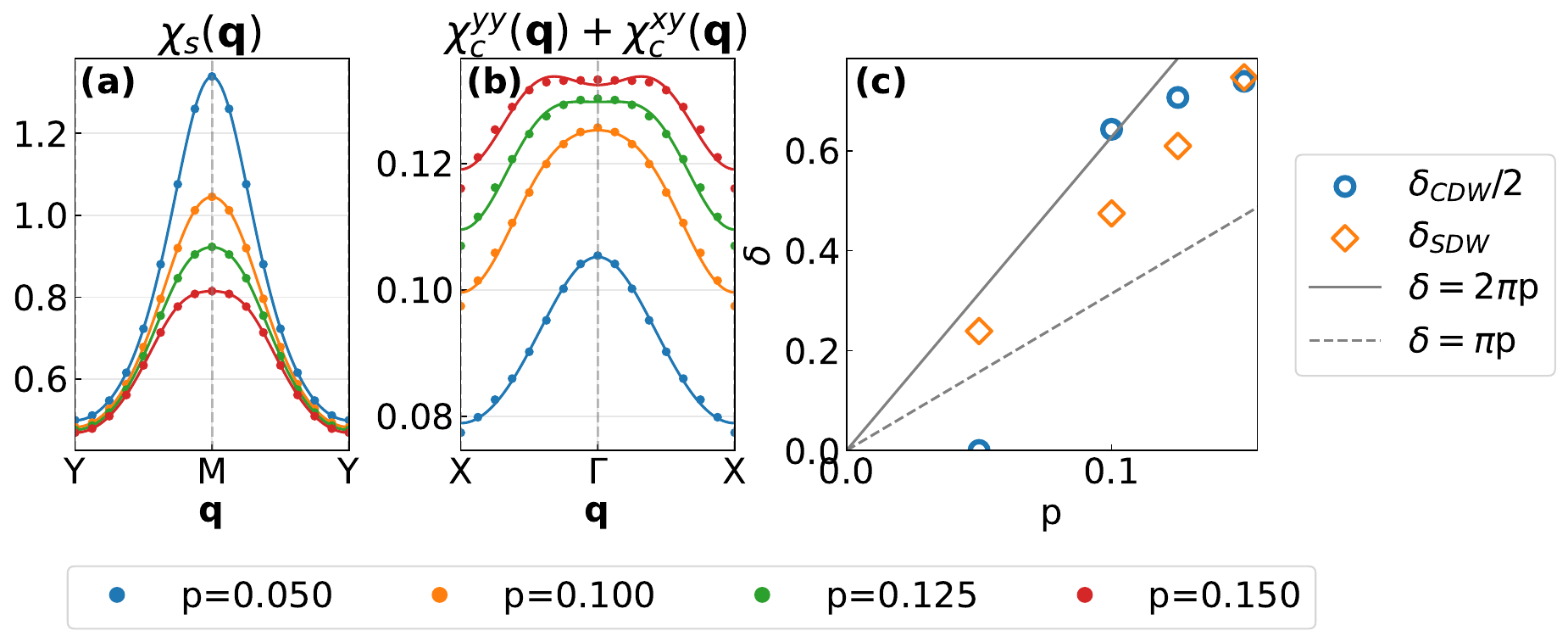}
    \caption{Same as Fig.~\ref{fig:Upp0Yamada} but for $U_{dd} = 8.5$~eV, $U_{pp} = 4.1$~eV, at $\beta = 6.5$~eV$^{-1}$.}
    \label{fig:Upp41Yamada}
\end{figure*}

\begin{figure*}
    \includegraphics[width=1.0\linewidth]{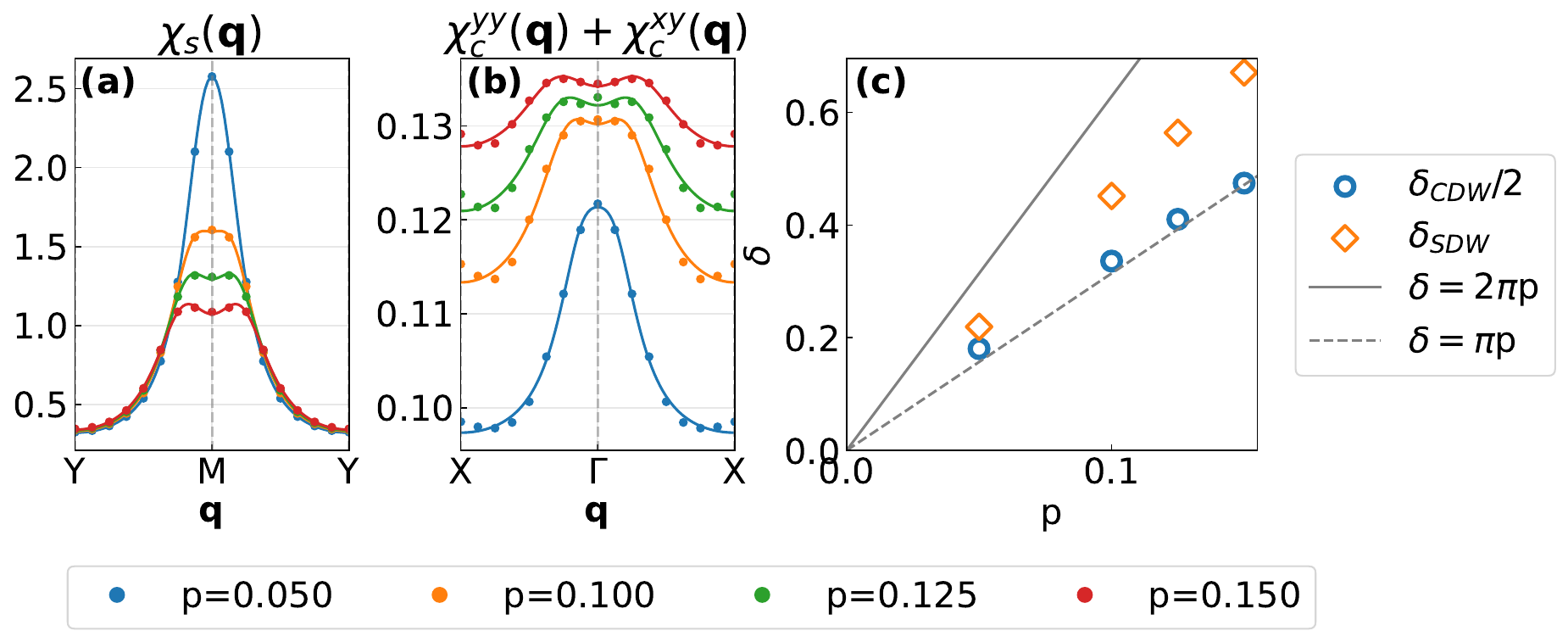}
    \caption{Same as Fig.~\ref{fig:Upp0Yamada} but for $U_{dd} = 6$~eV, $U_{pp} = 0$~eV, at $\beta = 12.0$~eV$^{-1}$.}
    \label{fig:Udd6Yamada}
\end{figure*}

\begin{figure*}
    \centering
    \includegraphics[width=1.0\linewidth]{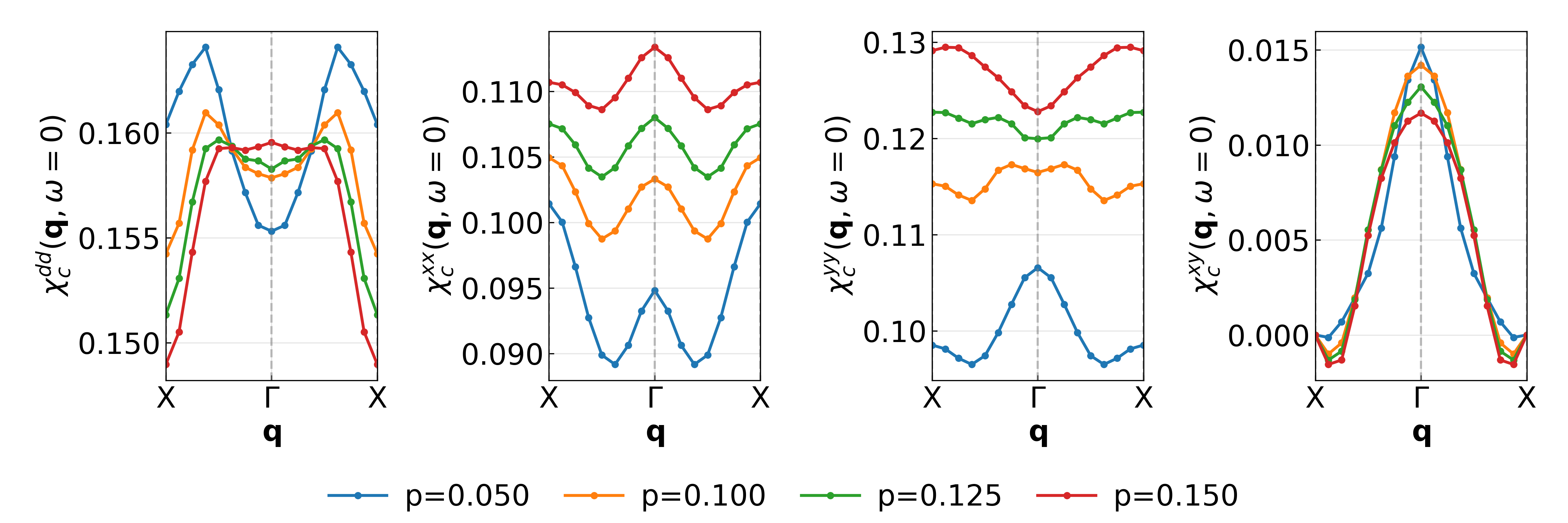}
    \caption{\textbf{Orbital resolved momentum space charge correlation $U_{dd} = 6$ eV $\beta = 12$ eV$^{-1}$}}
    \label{fig:nnk}
\end{figure*}

\begin{figure*}
    \centering
    \includegraphics[width=1.0\linewidth]{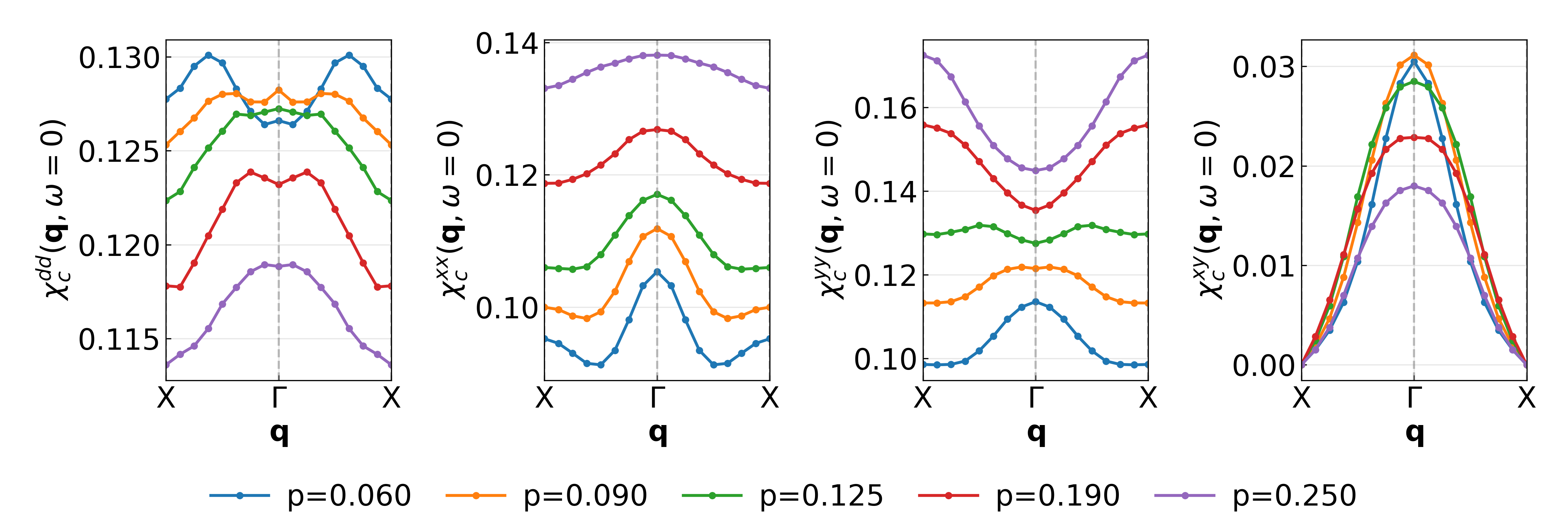}
    \caption{\textbf{Orbital resolved momentum space charge correlation $U_{pp} = 0$ $\beta = 10$ eV$^{-1}$}}
    \label{fig:nnk-Upp0}
\end{figure*}

\begin{figure*}
    \centering
    \includegraphics[width=1.0\linewidth]{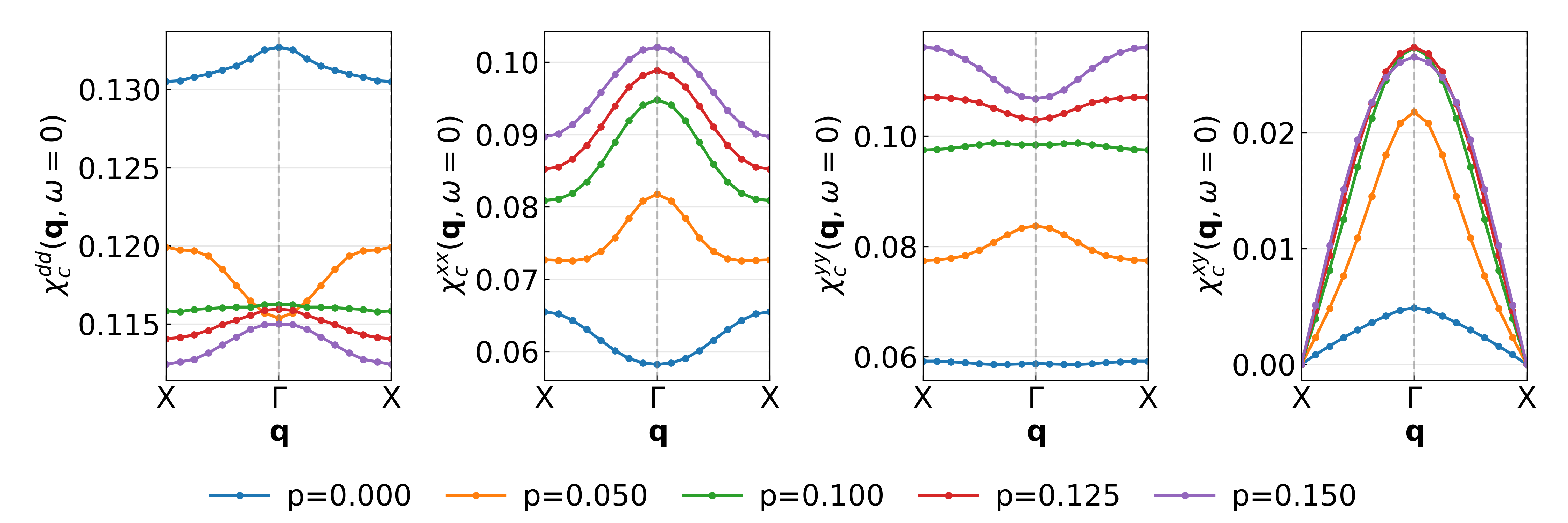}
    \caption{\textbf{Orbital resolved momentum space charge correlation $U_{pp} = 4.1$ eV $\beta = 6.5$ eV$^{-1}$}}
    \label{fig:nnk-Upp41}
\end{figure*}

\begin{figure*}
    \centering
    \includegraphics[width=1.0\linewidth]{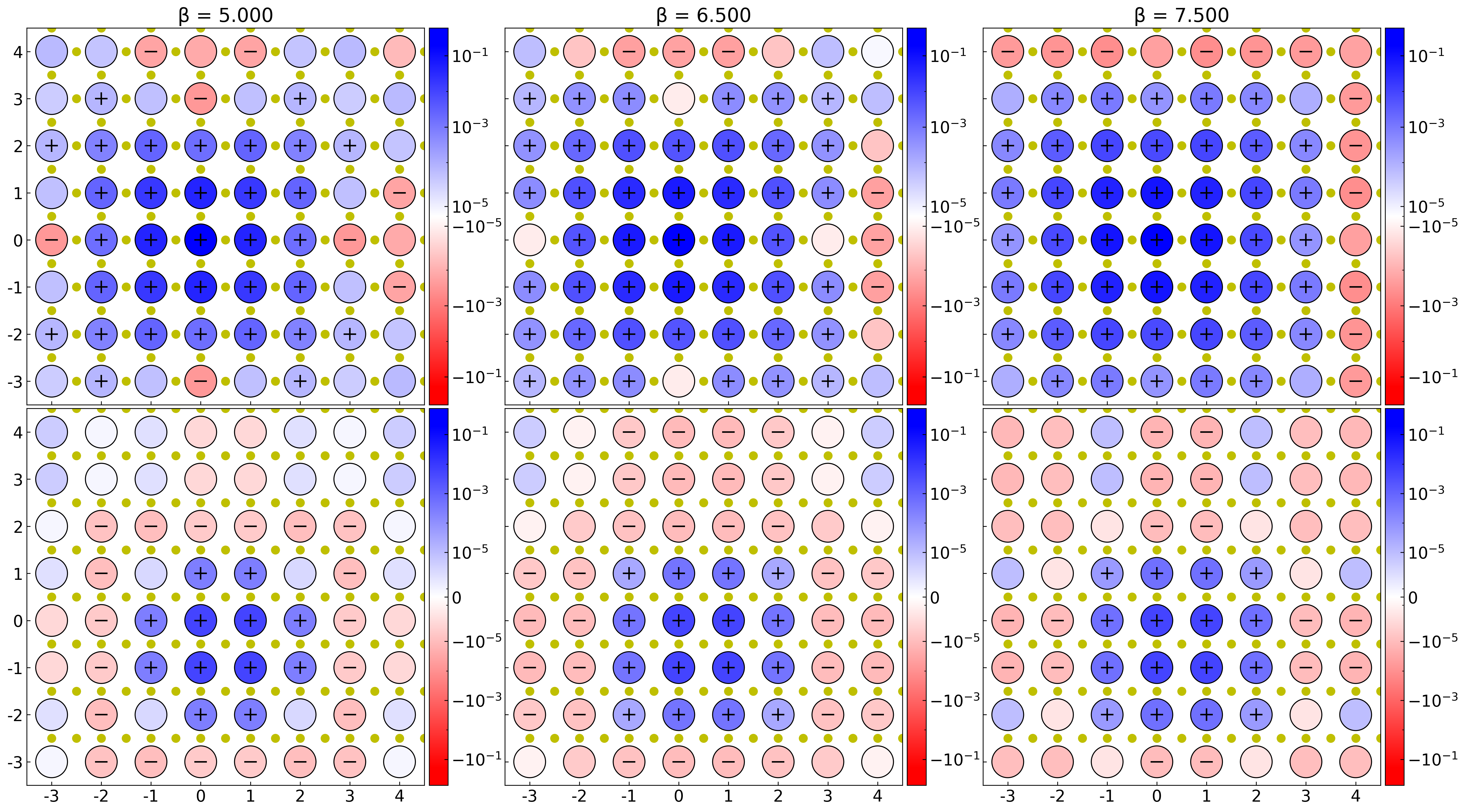}
    \caption{\textbf{Temperature evolution of staggered spin-spin and connected $p$-orbital charge-charge correlation} Same as Fig.~\ref{fig:temperature-locking} but for $U_{dd} = 8.5$ eV, $U_{pp} = 4.1$ eV, $\Delta_{pd} = 3.24$ eV at $p = 0.10$. As a general trend in the Emery model, the negatively correlated region of charge fluctuations moves towards the diagonal direction of the reference point as temperature decreases, indicating the development of $B_{1g}$ nematicity.}
    \label{fig:temperature-locking-U-n1.10_xy}
\end{figure*}

\begin{figure*}
    \centering
    \includegraphics[width=1.0\linewidth]{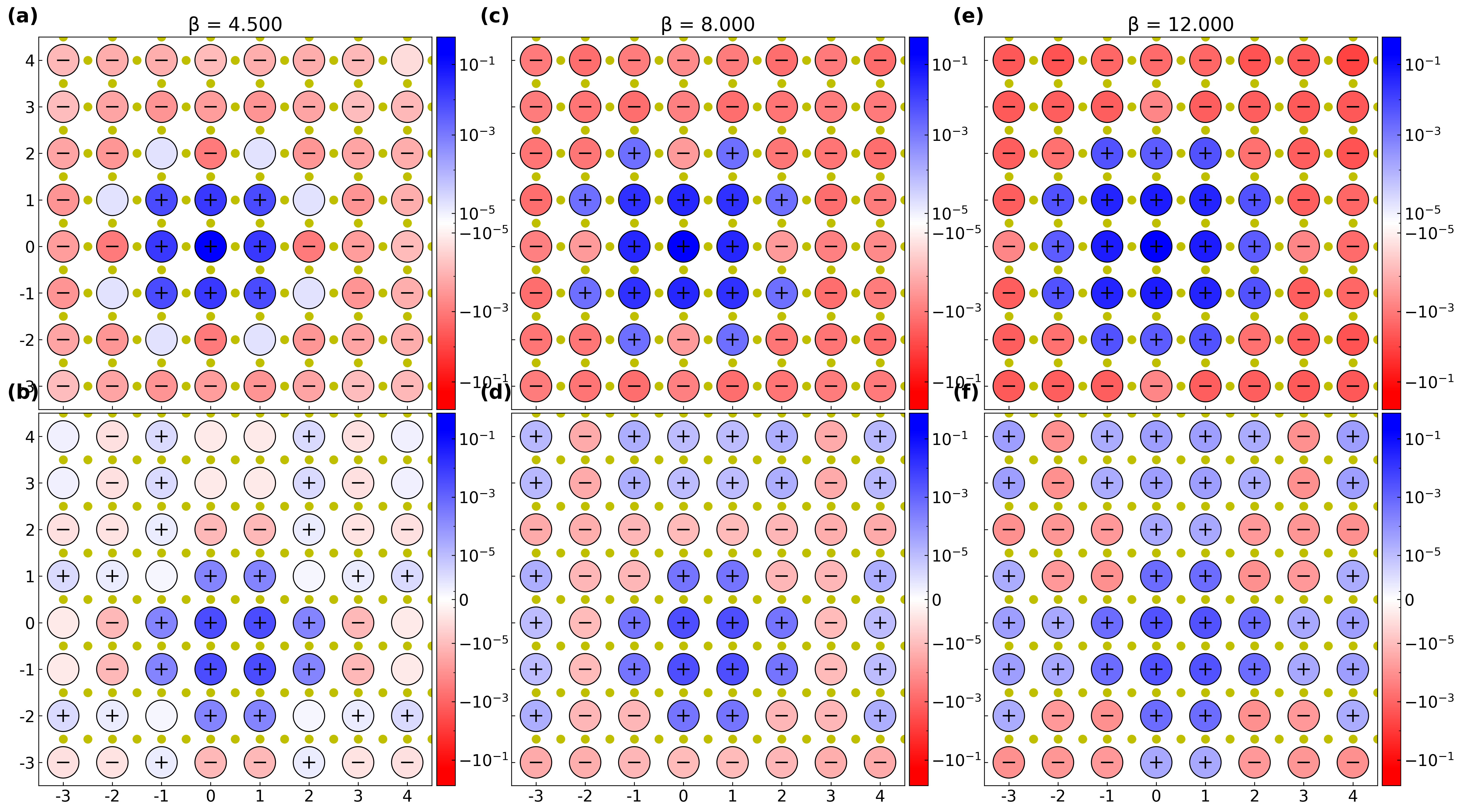}
    \caption{\textbf{Temperature evolution of staggered spin-spin and connected $p$-orbital charge-charge correlation} Same as Fig.~\ref{fig:temperature-locking} but at $p = 0.20$. As a general trend in the Emery model, the negatively correlated region of charge fluctuations moves towards the diagonal direction of the reference point as temperature decreases, indicating the development of axial nematicity.}
    \label{fig:temperature-locking-n1.20_xy}
\end{figure*}

\begin{figure*}
    \centering
    \includegraphics[width=1.0\linewidth]{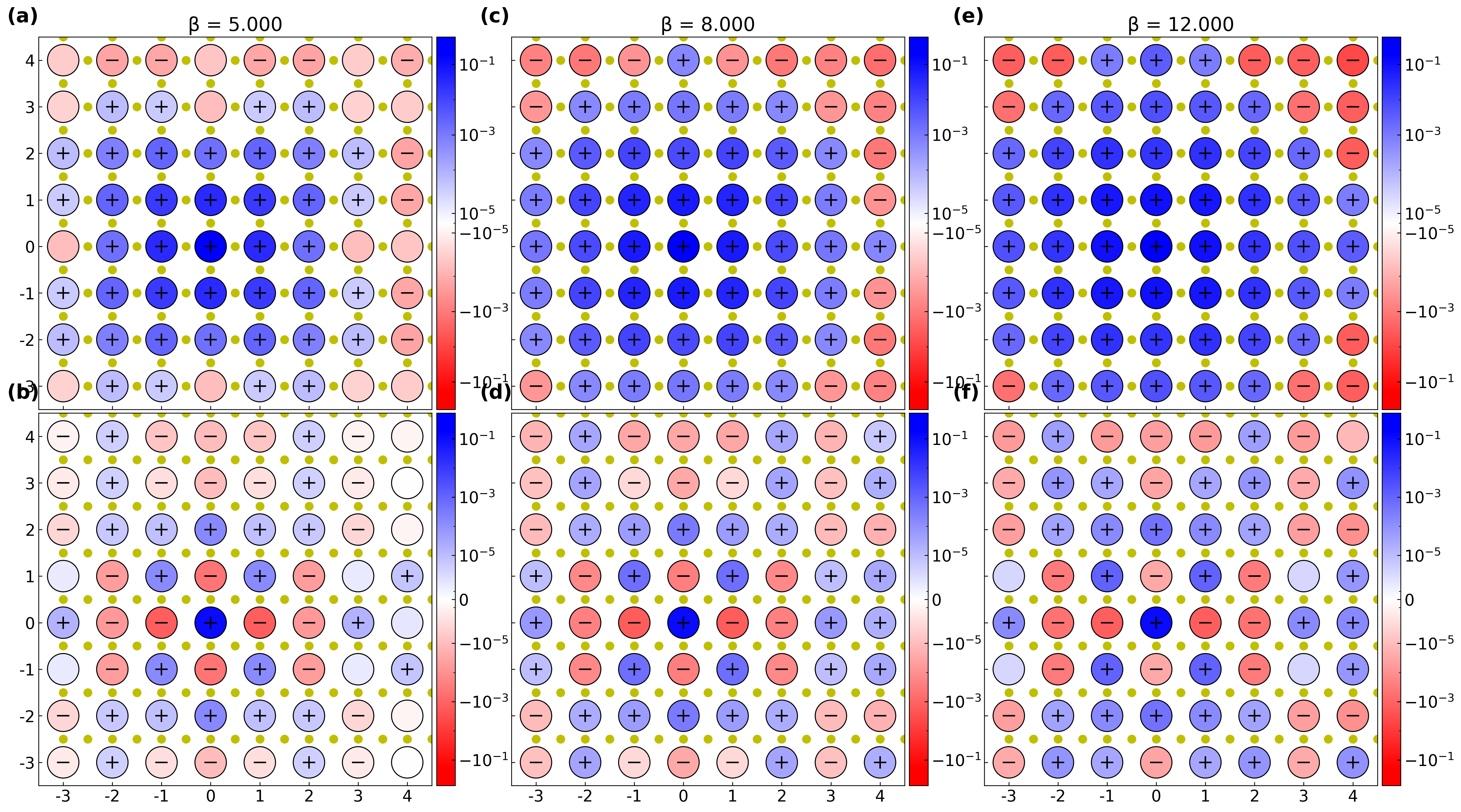}
    \caption{\textbf{Temperature evolution of staggered spin-spin and connected $p$-orbital charge-charge correlation} Same as Fig.~\ref{fig:temperature-locking} but for charge correlation between $p_y$ orbitals}
    \label{fig:temperature-locking_yy}
\end{figure*}

\begin{figure*}
    \centering
    \includegraphics[width=1.0\linewidth]{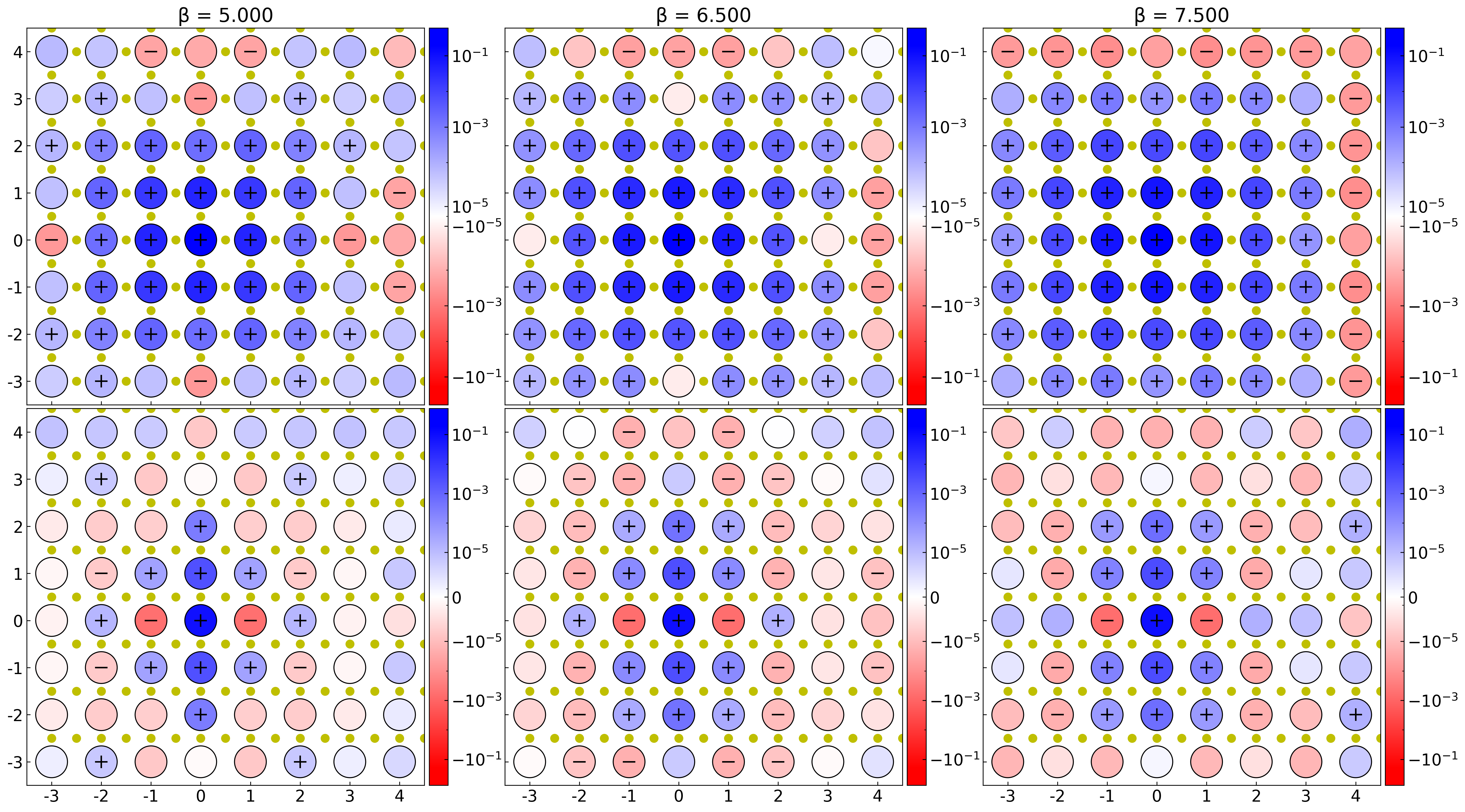}
    \caption{\textbf{Temperature evolution of staggered spin-spin and connected $p$-orbital charge-charge correlation} Same as Fig.~\ref{fig:temperature-locking-U-n1.10_xy} but for charge correlation between $p_y$ orbitals}
    \label{fig:temperature-locking-U-n1.10_yy}
\end{figure*}

\begin{figure*}
    \centering
    \includegraphics[width=1.0\linewidth]{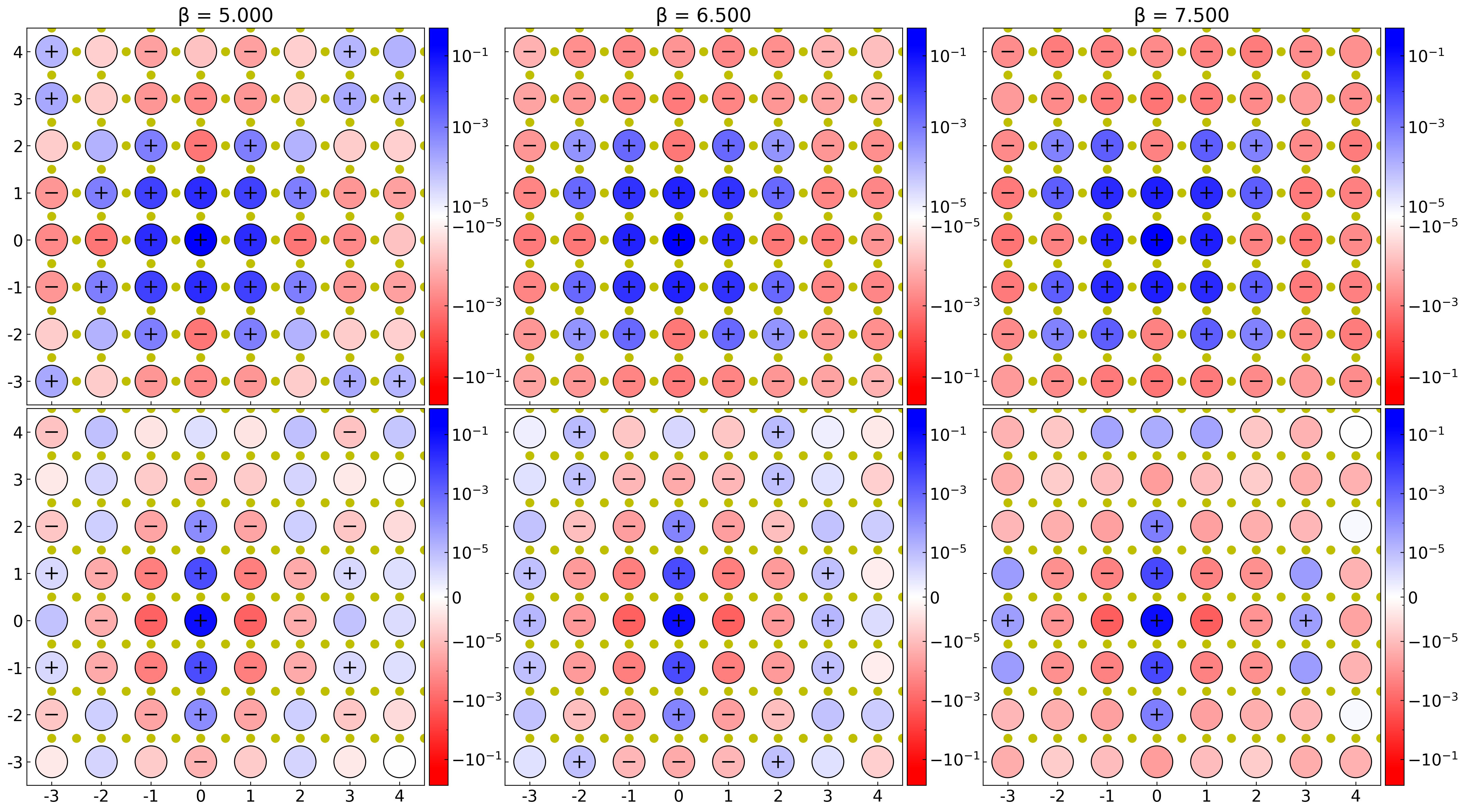}
    \caption{\textbf{Temperature evolution of staggered spin-spin and connected $p$-orbital charge-charge correlation} Same as Fig.~\ref{fig:temperature-locking-U-n1.10_yy} but for $p = 0.15$}
    \label{fig:temperature-locking-U-n1.15_yy}
\end{figure*}

\begin{figure*}
    \centering
    \includegraphics[width=0.7\linewidth]{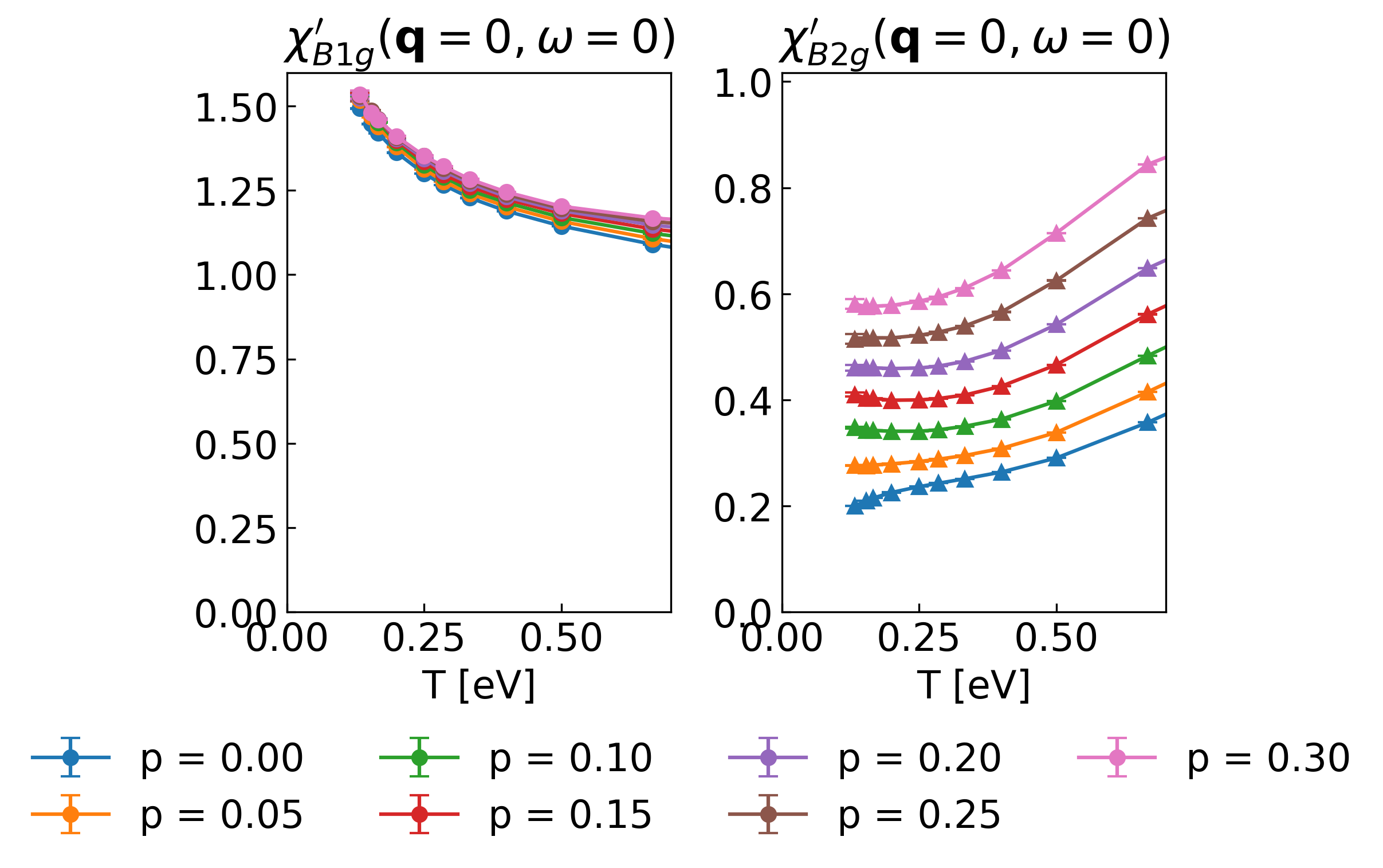}
    \caption{\textbf{Temperature evolution of bond-charge nematic susceptibilities} (a) Nearest neighbor Cu-O $B_{1g}$ susceptibility (b) Nearest neighbor O-O $B_{2g}$ susceptibility for $U_{dd} = 8.5~$eV and $U_{pp} = 4.1$~eV. (a) Anisotropy in kinetic energy induced by a small perturbation in $t_{pd}$ in $x$ and $y$ direction increases with decreasing temperature. $\chi_{B_{1g}}$ grows faster below $T\simeq 0.4$~eV. (b) Anisotropy in kinetic energy induced by a small perturbation in $t_{pp}$ decreases with decreasing temperature. $\chi_{B_{2g}}$ decreases with temperature faster above the $T\sim0.4$ eV, which is likely due to the charge transfer energy, and further decreases below $T\sim 0.1$ eV, echoing the temperature evolution of intertwined stripes in Fig.~\ref{fig:temperature-locking-U-n1.10_xy}.}
    \label{fig:rho-and-E-U}
\end{figure*}

\end{widetext}
\end{document}